\documentclass[pra, 10pt, twocolumn,letter, floatfix]{revtex4}
\usepackage{amsmath}
\usepackage{amsfonts}
\usepackage{amssymb}
\usepackage[T1]{fontenc}
\usepackage{graphicx}
\usepackage{rotating}
\allowdisplaybreaks
\usepackage{setspace}
\usepackage{bm}

\begin{document}

\title{Improved Optimization for the Cluster Jastrow Antisymmetric Geminal Power and Tests on Triple-Bond Dissociations}

\author{Eric Neuscamman$^{1,2,}$\footnote[1]{Electronic mail: eneuscamman@berkeley.edu}}
\affiliation{$^1$Department of Chemistry, University of California, Berkeley, California 94720, USA\\
             $^2$Chemical Sciences Division, Lawrence Berkeley National Laboratory, Berkeley, California 94720, USA}

\date{\today}

\begin{abstract}
We present a novel specialization of the variational Monte Carlo linear method for the optimization
of the recently introduced cluster Jastrow antisymmetric geminal power ansatz,
achieving a lower-order polynomial cost scaling
than would be possible with a naive application of the linear method and greatly improving
optimization performance relative to the previously employed quasi-Newton approach.
We test the methodology on highly multi-reference triple-bond stretches, achieving accuracies
superior to traditional coupled cluster theory and multi-reference perturbation theory
in both the typical example of N$_2$ and the transition-metal-oxide example of [ScO]$^{+}$.
\end{abstract}

\maketitle

\section{Introduction}
\label{sec:introduction}

One of the most pressing problems in quantum chemistry today is the challenge of predicting the detailed effects of electron correlation
in systems far from the mean-field regime such as molecules with stretched bonds, transition metal oxide catalysts,
and $\pi$-conjugated molecules with low-lying doubly-excited states.
While traditional quantum chemistry methods that build up from a Hartree-Fock reference function are very effective at describing
weak electron correlation (i.e.\ correlation that does not greatly alter the mean-field picture), and recent advances in
density matrix renormalization group (DMRG) \cite{Chan:2011:dmrg_in_chem} and full configuration interaction quantum Monte Carlo (FCI-QMC)
\cite{Booth:2015:mcscf_fciqmc}
have greatly expanded the reach of active space approaches to strong correlation (i.e.\ correlation, typically within the valence electrons,
that causes qualitatively non-mean-field effects), it remains difficult to affordably and accurately describe both weak and strong
correlation simultaneously.
Recently, we introduced \cite{Neuscamman:2013:cjagp} the cluster Jastrow antisymmetric geminal power (CJAGP) ansatz as a candidate to
address this challenge by attempting to combine the strengths of cluster operators \cite{BARTLETT:2007:cc_review},
Hilbert space Jastrow factors \cite{Neuscamman:2013:jagp}, and pairing wave functions, but we were limited in our ability to test this
new ansatz by the difficulty of combining quasi-Newton optimization techniques with variational Monte Carlo (VMC).
In this paper, we present a more robust and efficient optimization scheme for the CJAGP based on the VMC linear method (LM)
\cite{Nightingale:2001:linear_method,UmrTouFilSorHen-PRL-07,TouUmr-JCP-07,TouUmr-JCP-08} and use it to test this new ansatz
on two challenging triple-bond dissociations that were inaccessible to the old optimization method.

The ability of the CJAGP to encode strong correlation arises from its Jastrow-modified geminal power reference \cite{Neuscamman:2015:subtractive_jagp},
and so in a sense the theory can be seen as being part of the chemistry community's larger effort to construct ansatzes based on electron pairs.
Indeed, the ubiquity of electron pairing in molecular physics has spurred the investigation of numerous pair-based approaches to electron correlation,
in which the fundamental wave function building block is a two-electron geminal rather than a one-electron orbital.
Early examples include perfect pairing (PP) \cite{POPLE:1953:agp,Beran:2005:upp},
the ``bare'' (i.e.\ not Jastrow-modified) antisymmetric geminal power (AGP) \cite{Bratoz:1965:AGP,COLEMAN:1965:AGP,Scuseria:2002:hfb},
and products of strongly orthogonal geminals \cite{Kutzelnigg:1964:apsg,Kutzelnigg:1965:apsg,Surjan:2012:apsg}.
More recently, there has been renewed interest in pairing wave functions based on the idea of relaxing the strong orthogonality constraint,
as in generalizations of PP \cite{Head-Gordon:2000:non_orth_pp,Head-Gordon:2000:imperfect_pairing,Head-Gordon:2002:gvb_cc}
the antisymmetric product of 1-reference-orbital geminals (AP1roG)
\cite{Bultinck:2013:nonorth_gems,Ayers:2014:nonvar_oo_ap1rog,VanNeck:2014:ap1rog_on_hubbard,VanNeck:2014:seniority_2_ap1rog_oo,Ayers:2014:geminal_accuracy,Ayers:2015:ap1rog_lcc}
and extensions of the singlet-type strongly orthogonal geminal (SSG) approach
\cite{Rassolov:2002:ssg,Rassolov:2004:pert_ssg,Rassolov:2007:ssg,Rassolov:2007:spin_proj_ssg,Rassolov:2014:sspg,Szabados:2015:spin_proj_ssg}.
While the CJAGP has strong connections to these pairing theories, it is important to recognize that Jastrow-modification can
drastically change the ansatz, and it is actually the combination of Jastrow factor and geminal power that lies at the heart of the ansatz's ability
to capture strong correlation \cite{Neuscamman:2015:subtractive_jagp}.
For this reason, the pairing theory that most closely relates to CJAGP is JAGP with real space Jastrows
\cite{Sorella:2003:agp_sr,Sorella:2004:agp_sr,Sorella:2007:jagp_vdw,Sorella:2009:jagp_molec}, although we must emphasize that
real space and Hilbert space Jastrow factors are quite different, and so many of the approximations involved are distinct.

The ability of the CJAGP to encode weak correlation arises from the fact that under a unitary orbital rotation, the Hilbert space Jastrow factor
becomes a simplified coupled cluster (CC) doubles operator \cite{Neuscamman:2013:cjagp} similar in structure to the tensor hypercontraction representation
of doubles amplitudes \cite{Martinez:2012:thc_correlated}.
The variational freedom of the cluster-Jastrow (CJ) operator is much reduced compared to the traditional CC doubles operator
\cite{BARTLETT:2007:cc_review}, and as we will discuss below this simplicity may limit the CJAGP's ability to encode the finer details of dynamic correlation.
Note that the CJ operator is \textit{not} a pairing operator, and that the electron pairing qualities of CJAGP come instead from its AGP reference.
One must therefore be careful not to confuse the CJ operator with the CC operator representations of various pairing theories,
such as 
PP \cite{Ukrainskii:1977,Cullen:1996:gvb_from_cc}, some forms of the generalized valence bond \cite{Head-Gordon:2001:gvb_cc},
AP1roG \cite{Ayers:2014:nonvar_oo_ap1rog,VanNeck:2014:ap1rog_on_hubbard,VanNeck:2014:seniority_2_ap1rog_oo},
and pair CC doubles \cite{Scuseria:2014:sen_0_pair_ccd,Henderson:2014:pair_ccd_attractive,Scuseria:2014:seniority_cc}.
Indeed, these theories often use their pairing ansatzes' cluster operator formulation to facilitate a non-variational, projective optimization scheme
as in traditional CC theory, whereas CJAGP is evaluated using \textit{variational} Monte Carlo.
As such, it may be conceptually more useful to see CJAGP as an attempt to achieve a type of variational, multi-reference CC, inspired
by the accuracy seen in studies of variational and quasi-variational CC
\cite{Head-Gordon:2000:var_cc,Knowles:2010:vcc,Knowles:2012:quasi_var_cc,Knowles:2012:qvcc_benchmark,Knowles:2012:qvcc_nonlin_optical,Knowles:2012:qvcc_pert_triples}
and the extraordinary accuracies achievable by multi-reference CC \cite{PaldusLi:1999:cc_review}.

The remainder of this paper is organized as follows.
We begin by defining the CJAGP ansatz (Section \ref{sec:basics}) and reviewing the typical formulation of the LM (Section \ref{sec:tlm}).
We then show how the cost-scaling for applying the LM to the CJAGP may be reduced (Section \ref{sec:lsmb}), how the
strong zero variance principle is maintained (Section \ref{sec:szv}), and how one can avoid constructing the LM matrices
when desirable (Section \ref{sec:amb}).
After presenting computational details (Section \ref{sec:comp_det}), we then present data on the improved optimization
efficiency (Section \ref{sec:convergence}) as well as the accuracy of the method in the triple bond dissociations of N$_2$ (Section \ref{sec:n2})
and [ScO]$^+$ (Section \ref{sec:sco}), before concluding and offering remarks on possible future directions (Section \ref{sec:conclusions}).

\section{Theory}
\label{sec:theory}

\subsection{Basics}
\label{sec:basics}

In this paper we seek to optimize the CJAGP ansatz,
\begin{align}
|\Psi\rangle = \exp(\hat{\mathcal{K}}) |\Phi\rangle,
\label{eqn:cjagp}
\end{align}
in which the unitary orbital rotation operator $\exp(\hat{\mathcal{K}})$ is defined by the
anti-Hermitian operator
\begin{align}
\hat{\mathcal{K}} = \sum_{p<q} K_{pq} ( a^+_p a_q - a^+_q a_p )
\label{eqn:k}
\end{align}
and
\begin{align}
|\Phi\rangle = \exp \left( \sum_{ij} J_{ij} \hat{n}_i \hat{n}_j \right)
\left( \sum_{rs} F_{rs} a^+_r a^+_s \right)^{N/2}
|0\rangle
\label{eqn:jagp}
\end{align}
is the JAGP ansatz with pairing matrix $\bm{F}$ and Jastrow factor coefficients $\bm{J}$.
In Eq.\ (\ref{eqn:jagp}), $N$ is the (even) number of electrons, $r$ and $s$ are restricted
to $\alpha$ and $\beta$ spin-orbitals, respectively, and $i$ and $j$ range over all spin-orbitals.
Note that unless otherwise stated, indices in this paper are assumed to range over all spin-orbitals.
We will make use of the fermionic creation and destruction operators,
$a^+_p$ and $a_p$, which create or destroy an electron in spin-orbital $p$ and which
obey the usual anti-commutation rules.
We also employ the number operators $\hat{n}_p = a^+_p a_p$.

The development of improved optimization methods for the orbital rotation defined by $\hat{\mathcal{K}}$
is important because it is this rotation that allows the Jastrow factor to act as a limited
CC doubles operator,
\begin{align}
e^{\hat{\mathcal{K}}} e^{\sum_{ij} J_{ij} \hat{n}_i \hat{n}_j} e^{-\hat{\mathcal{K}}}
& = \exp \left( \sum_{ijkl} T_{ij}^{kl} a^+_k a_i a^+_l a_j \right),
\label{eqn:cluster_op} \\
T_{ij}^{kl}
& = \sum_{pq} U^*_{ip} U_{kp} J_{pq} U^*_{jq} U_{lq},
\label{eqn:cluster_amp}
\end{align}
where $\bm{U}$ results from exponentiating the antisymmetrization of the upper-triangular $\bm{K}$ \cite{Neuscamman:2013:cjagp}.
Given the potentially highly multi-reference nature of the geminal power \cite{Neuscamman:2015:subtractive_jagp},
this raises the tantalizing question of whether the CJAGP can act as an effective surrogate for
much more complex complete-active-space-based multireference CC ansatzes that have outstanding
accuracy but steeply scaling computational costs.
Although initial investigations into the CJAGP
showed promise \cite{Neuscamman:2013:cjagp}, they were limited by the shortcomings
of combining the quasi-Newton L-BFGS method with VMC.
We will therefore turn our attention to creating a more effective optimization scheme in order to
push CJAGP into larger and more interesting systems.

\subsection{Traditional Linear Method}
\label{sec:tlm}

The LM \cite{Nightingale:2001:linear_method,UmrTouFilSorHen-PRL-07,TouUmr-JCP-07,TouUmr-JCP-08}
optimization scheme
works by solving the Schr\"{o}dinger equation (SE) in the subspace of Hilbert space spanned by
the approximate wave function and its first derivatives with respect to its variables $\bm{\mu}$, which
we write concisely as
\begin{align}
|\Psi^0\rangle \equiv |\Psi\rangle \qquad |\Psi^x\rangle \equiv
\frac{\partial|\Psi\rangle}{\partial \mu_x} \quad x\in \{1,2,...,n_\mathrm{v}\}.
\label{eqn:psi_var_der}
\end{align}
As these functions may not be orthogonal, the SE to be solved is a generalized eigenvalue problem,
\begin{alignat}{2}
\bm{H} \bm{c} &= E \bm{S} \bm{c}
\label{eqn:lm_se} \\
H_{xy} &= \langle\Psi^x|\hat{H}|\Psi^y\rangle && \quad \forall \quad x,y\in\{0,1,2,...,n_\mathrm{v}\} \label{eqn:lm_h} \\
S_{xy} &= \langle\Psi^x|\Psi^y\rangle         && \quad \forall \quad x,y\in\{0,1,2,...,n_\mathrm{v}\} \label{eqn:lm_s}
\end{alignat}
Assuming the initial wave function is close to the energy minimum, then the ratios $c_x/c_0$ for $x>0$ can be expected to be
small, as the optimal wave function in the LM subspace should be a small change from $|\Psi\rangle$ (this smallness
can be ensured by penalizing the $x>0$ diagonal elements $H_{xx}$ \cite{TouUmr-JCP-07}).
Having solved Eq.\ (\ref{eqn:lm_se}) for $\bm{c}$, we may then update our wave function by a reverse Taylor expansion,
\begin{align}
|\Psi(\bm{\mu})\rangle \rightarrow |\Psi(\bm{\mu} + \text{\reflectbox{$\bm{c}$}}/c_0)\rangle \approx |\Psi\rangle + \sum_{x=1}^{n_\mathrm{v}} \frac{c_x}{c_0}|\Psi^x\rangle ,
\label{eqn:reverse_taylor}
\end{align}
where \text{\reflectbox{$\bm{c}$}} is the length-$n_\mathrm{v}$ vector obtained by removing the first element ($c_0$) from $\bm{c}$.
The key role of Monte Carlo is to evaluate the matrices $\bm{H}$ and $\bm{S}$, which is done by a resolution of the identity in terms
of occupation number vectors $\bm{n}$ (in real space we would instead use an integral over positions) over which
a stochastic sample is taken,
\begin{align}
A_{xy} &= \sum_{\bm{n}} \langle\Psi^x|\bm{n}\rangle \langle\bm{n}|\hat{A}|\Psi^y\rangle
\notag \\
       &= \sum_{\bm{n}} |\langle\bm{n}|\Psi\rangle|^2 \frac{\langle\Psi^x|\bm{n}\rangle}{\langle\Psi|\bm{n}\rangle}
                                                      \frac{\langle\bm{n}|\hat{A}|\Psi^y\rangle}{\langle\bm{n}|\Psi\rangle}
\notag \\
       &\approx \sum_{\bm{n}\in\xi} \frac{\langle\Psi^x|\bm{n}\rangle}{\langle\Psi|\bm{n}\rangle}
                                    \frac{\langle\bm{n}|\hat{A}|\Psi^y\rangle}{\langle\bm{n}|\Psi\rangle}
\label{eqn:trad_mc_mat}
\end{align}
For $\bm{H}$ we set $\hat{A}=\hat{H}$ while for $\bm{S}$ we set $\hat{A}$ to the identity operator.
In this paper the sample of configurations $\xi$ will be drawn from the distribution $|\langle\bm{n}|\Psi\rangle|^2$, but any
distribution $|Q(\bm{n})|^2$ can be used if the right hand side of Eq.\ (\ref{eqn:trad_mc_mat}) is modified to
$\sum_{\bm{n}\in\xi}\langle\Psi^x|\bm{n}\rangle\langle\bm{n}|\hat{A}|\Psi^y\rangle/|Q(\bm{n})|^2$.
Note that the normalization constant for the sampled distribution may be ignored, as it will appear on either side of
Eq.\ (\ref{eqn:lm_se}) and will thus not affect the solution $\bm{c}$.
For CJAGP, we will retain the use of Eq.\ (\ref{eqn:trad_mc_mat}) for most but not all elements of $\bm{H}$ and $\bm{S}$,
as shown in Figure \ref{fig:matrix_tiles}.

To see why we do not retain the traditional approach for all matrix elements, consider element $H_{xy}$ in which $\mu_y$
is the orbital rotation variable $K_{pq}$, in which case we must evaluate
\begin{align}
\langle\bm{n}|\hat{H}|\Psi^y\rangle
=  \frac{\partial \langle\bm{n}|\hat{H}|\Psi\rangle}{\partial K_{pq}} 
= \langle\bm{n}|\hat{H} ( a^{+}_p a_q - a^{+}_q a_p ) |\Psi\rangle,
\label{eqn:hard_term}
\end{align}
in which the two-electron component of $\hat{H}$ combines with the $pq$-indexed excitations
to create triple excitations acting on the configuration $\bm{n}$.
While such triple-excitation terms may be evaluated using the same approach as for double excitations (as in the JAGP energy evaluation \cite{Neuscamman:2013:jagp}),
the cost scaling for this approach is $N^6$, which is much higher than the $N^4$ scaling that can be achieved \cite{Neuscamman:2013:jagp} when $\mu_y$
corresponds to a Jastrow or AGP variable.
(Note that to get the LM's overall cost scaling, one must add an additional factor of $N$ if the statistical uncertainty of extensive quantities is to be held constant
due to the requisite increase in the sample length.)

\subsection{Lower Scaling Matrix Builds}
\label{sec:lsmb}

For a general two-body operator of the form
\begin{align}
\hat{A} = A_0 + \sum_{pq} A^p_q a^+_p a_q + \sum_{pqrs} A^{pq}_{rs} a^+_p a^+_q a_s a_r
\label{eqn:gen_2body_op}
\end{align}
and a wave function ansatz consisting of a JAGP augmented by an
orbital rotation as in Eq.\ (\ref{eqn:cjagp}),
the per-sample cost scaling to build the matrix $\bm{A}$ can be reduced to
$N^5$ by working in the one-particle basis in which $\hat{\mathcal{K}}=0$
and by performing the Monte-Carlo-sampled resolution of the identity in a
slightly different way.
Note that an arbitrary rotation of the one-particle basis (after which $\hat{A}$
will have the same form but different coefficients) can be achieved by converting
\begin{align}
|\Psi\rangle \rightarrow e^{-\hat{\mathcal{L}}}|\Psi\rangle
\qquad
\hat{A} \rightarrow e^{-\hat{\mathcal{L}}}\hat{A}e^{\hat{\mathcal{L}}}
\label{eqn:rotate_basis}
\end{align}
using an anti-Hermitian one-body operator $\hat{\mathcal{L}}$ that defines the rotation.
At the end of each LM iteration, at which point $\hat{\mathcal{K}}$ may be nonzero
due to the LM update of Eq.\ (\ref{eqn:reverse_taylor}), we may thus ``reset'' $\hat{\mathcal{K}}$ to 0
via a basis-rotation with $\hat{\mathcal{L}}=\hat{\mathcal{K}}$.
The one- and two-electron coefficients needed to represent $\hat{A}$ in the new basis,
i.e.\ $A^p_q$ and $A^{pq}_{rs}$ in Eq.\ (\ref{eqn:gen_2body_op}),
can be evaluated at an $N^5$ cost as per a standard atomic-to-molecular-orbital conversion
of the one- and two-electron integrals \cite{Helgaker_book}.
As the basis rotation is required only once per LM iteration, rather than once per sample, its
cost is negligible compared to the sampling effort involved in estimating the matrix $\bm{A}$.

\begin{figure}[t]
\centering
\includegraphics[width=7.5cm,angle=0]{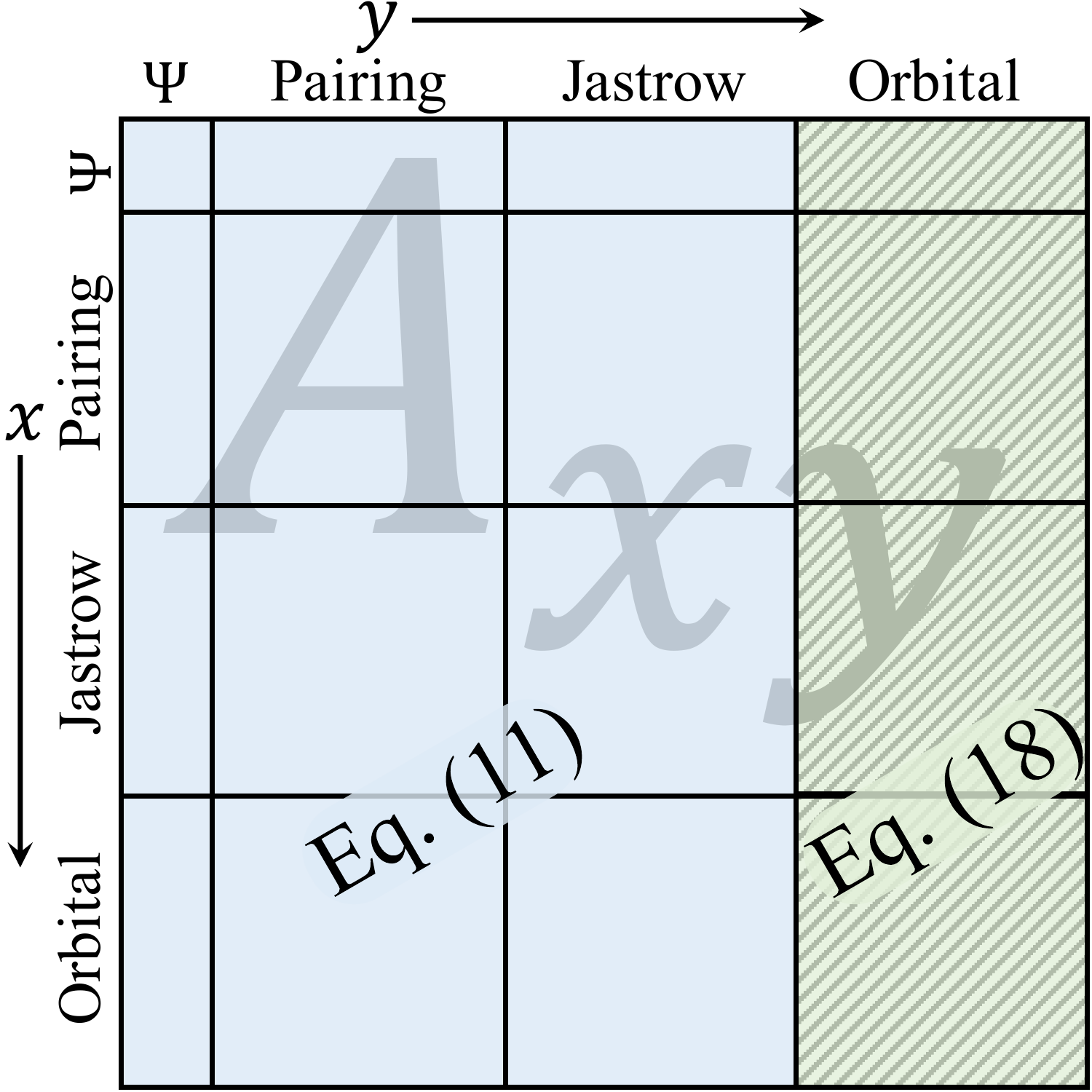}
\caption{Equations used for evaluating different subsections of the LM matrices $\bm{H}$ and $\bm{S}$.
        }
\label{fig:matrix_tiles}
\end{figure}

Working in the $\hat{\mathcal{K}}=0$ one-particle basis, we may express the difficult $\mu_y=K_{pq}$ matrix element as
\begin{align}
A_{xy} & = \langle\Psi^x|\hat{A}|\Psi^y\rangle
\notag \\
       & = \left[ \langle\Psi^x| \frac{\partial}{\partial K_{pq}} \left( \hat{A} e^{\hat{\mathcal{K}}} |\Phi\rangle \right) \right]_{\hat{\mathcal{K}}=0}
\notag \\
       & = \left[ \langle\Psi^x| \frac{\partial}{\partial K_{pq}} \left( e^{\hat{\mathcal{K}}} e^{-\hat{\mathcal{K}}} \hat{A} e^{\hat{\mathcal{K}}} |\Phi\rangle \right) \right]_{\hat{\mathcal{K}}=0}
\notag \\
       & = \langle\Psi^x| \hat{C} |\Phi\rangle + \langle\Psi^x| \hat{D} \hat{A} |\Phi\rangle
\notag \\
       & = \sum_{\bm{n}}   \langle\Psi^x|\bm{n}\rangle \langle\bm{n}|\hat{C}|\Phi\rangle
                         + \langle\Psi^x|\hat{D}|\bm{n}\rangle \langle\bm{n}|\hat{A}|\Phi\rangle
\label{eqn:double_ri}
\end{align}
where we have defined
\begin{align}
\hat{C} & \equiv \left[ \frac{\partial (e^{-\hat{\mathcal{K}}} \hat{A} e^{\hat{\mathcal{K}}})}{\partial K_{pq}} \right]_{\hat{\mathcal{K}}=0} = \left[ \hat{A}, \hspace{1mm} a^+_p a_q - a^+_q a_p \right]
\label{eqn:der_of_eKAeK_wrt_K} \\
\hat{D} & \equiv \left[ \frac{\partial e^{\hat{\mathcal{K}}}}{\partial K_{pq}} \right]_{\hat{\mathcal{K}}=0} = a^+_p a_q - a^+_q a_p
\label{eqn:der_of_eK_wrt_K}
\end{align}
The rationale for these placements of the identity resolutions is that they
isolate the difficult operators $\hat{A}$ and $\hat{C}$ such that no term involves
more than a double excitation on $|\bm{n}\rangle$ (the uncontracted triple excitations in the
commutator of $\hat{C}$ cancel each other as usual), thus avoiding the triple excitation
in Eq.\ (\ref{eqn:hard_term}) that led to $N^6$ scaling.
Having placed our identity resolutions, we may now evaluate them
stochastically on a sample $\xi$ drawn from $|\langle\Phi|\bm{n}\rangle|^2$ in order
to produce our Monte Carlo estimate of the matrix element:
\begin{align}
A_{xy} & \approx \sum_{\bm{n}\in\xi} \frac{\langle\Psi^x|\bm{n}\rangle}{\langle\Phi|\bm{n}\rangle}
                                     \frac{\langle\bm{n}|\hat{C}|\Phi\rangle}{\langle\bm{n}|\Phi\rangle}
         +                           \frac{\langle\Psi^x|\hat{D}|\bm{n}\rangle}{\langle\Phi|\bm{n}\rangle}
                                     \frac{\langle\bm{n}|\hat{A}|\Phi\rangle}{\langle\bm{n}|\Phi\rangle}
\label{eqn:new_Axy}
\end{align}
It now remains to evaluate these matrix element estimates for the identity and Hamiltonian operators
involved in the LM.

For the overlap matrix $\bm{S}$, for which $\hat{A}$ is the identity, 
$\hat{C}$ vanishes and Eq.\ (\ref{eqn:new_Axy}) simplifies to
\begin{align}
S_{xy} & \approx \sum_{\bm{n}\in\xi} \frac{\langle\Psi^x|(a^+_p a_q - a^+_q a_p)|\bm{n}\rangle}{\langle\Phi|\bm{n}\rangle}.
\label{eqn:new_Sxy}
\end{align}
As shown in Appendix \ref{sec:oroc}, the per-sample cost to evaluate these $\mu_y=K_{pq}$ matrix blocks
(i.e. the Eq.\ (\ref{eqn:new_Axy}) blocks for $\bm{S}$ in Figure \ref{fig:matrix_tiles})
grows as only $N^4$.

For the Hamiltonian matrix $\bm{H}$, for which $\hat{A}=\hat{H}$, things are not so simple,
although it is possible to avoid an $N^6$ per-sample cost scaling.
To begin, we may recognize that the right hand part Eq.\ (\ref{eqn:new_Axy}) becomes a simple
modification of Eq.\ (\ref{eqn:new_Sxy}) in which each term is scaled by
the JAGP local energy $\langle\bm{n}|\hat{H}|\Phi\rangle / \langle\bm{n}|\Phi\rangle$ (which can be evaluated at an
$N^4$ per-sample cost \cite{Neuscamman:2013:jagp}),
and so its contribution to $\bm{H}$ can be evaluated at an $N^4$ per-sample cost by a direct analogue of
the approach for $\bm{S}$ given in Appendix \ref{sec:oroc}.
In the left-hand part of Eq.\ (\ref{eqn:new_Axy}), consider first the derivative ratios
\begin{align}
\mathcal{D}_{\bm{n}}(\mu_x) \equiv \frac{\langle\Psi^x|\bm{n}\rangle} {\langle\Phi|\bm{n}\rangle}.
\end{align}
For $\mu_x$ either a pairing matrix element or a Jastrow coefficient, these ratios have been evaluated
previously for the JAGP \cite{Neuscamman:2013:jagp}.
When $\mu_x$ is an orbital rotation variable $K_{pq}$, the ratios are
\begin{align}
\mathcal{D}_{\bm{n}}(K_{pq}) = \frac{\langle\Phi|(a^+_p a_q - a^+_q a_p)|\bm{n}\rangle} {\langle\Phi|\bm{n}\rangle},
\end{align}
which can be evaluated efficiently as shown in Appendix \ref{sec:oroc}, specifically in Eq.\ (\ref{eqn:pq_ratio}).

The final term needed to construct $\bm{H}$, and the one responsible for the overall $N^5$ per-sample cost scaling of the
construction, is the term in Eq.\ (\ref{eqn:new_Axy}) containing $\hat{C}$.
We will worry only about the two-electron component of $\hat{H}$ (the reader may convince herself that the one-electron
component is less expensive), for which we must evaluate
\begin{align}
\frac{1} {\langle\bm{n}|\Phi\rangle}
\langle\bm{n}| \left[ \hspace{1mm} \sum_{ijkl} \hspace{1mm} g^{ij}_{kl} \hspace{1mm} a^+_i a^+_j a_l a_k \hspace{1mm},
                      \hspace{1mm} a^+_p a_q - a^+_q a_p \hspace{0.7mm} \right] |\Phi\rangle
\label{eqn:two_elec_comm}
\end{align}
where $g^{ij}_{kl}$ are the usual two-electron integrals \cite{Helgaker_book}.
Defining the double excitation ratios
\begin{align}
Q^{ij}_{kl} \equiv \frac{\langle\bm{n}| a^+_i a^+_j a_l a_k |\Phi\rangle} {\langle\bm{n}|\Phi\rangle},
\label{eqn:double_excite_ratios}
\end{align}
which are derivatives of the JAGP local energy with respect to $g^{ij}_{kl}$ (see Eq.\ (34) of Ref.\ \cite{Neuscamman:2013:jagp}) and can
thus all be evaluated for the same $N^4$ cost-per-sample scaling as the local energy itself,
one may expand Eq.\ (\ref{eqn:two_elec_comm}) as
\begin{align}
\sum_{ i j k }
\Big( \hphantom{+}
   \hspace{1mm} & g^{ i j }_{ p k } Q^{ i j }_{ q k }
 +                g^{ i j }_{ k p } Q^{ i j }_{ k q }
 -                g^{ i q }_{ j k } Q^{ i p }_{ j k }
 -                g^{ q i }_{ j k } Q^{ p i }_{ j k } \notag \\
 + \hspace{1mm} & g^{ i j }_{ p k } Q^{ q k }_{ i j }
 +                g^{ i j }_{ k p } Q^{ k q }_{ i j }
 -                g^{ i q }_{ j k } Q^{ j k }_{ i p }
 -                g^{ q i }_{ j k } Q^{ j k }_{ p i }
\hspace{1.5mm} \Big).
\label{eqn:expanded_commutator}
\end{align}
Each of these terms can clearly be evaluated for a per-sample cost scaling as $N^5$,
giving the explicit construction of the CJAGP $\bm{H}$ matrix according to the scheme in
Figure \ref{fig:matrix_tiles} an overall per-sample cost that scales as $N^5$.
This is better than the $N^6$ per-sample cost resulting from a naive application of the traditional
LM matrix build, but nonetheless a higher scaling than for JAGP.

\subsection{Strong Zero Variance}
\label{sec:szv}

In the traditional LM, the stochastic approximation to the generalized eigenvalue problem in Eq.\ (\ref{eqn:lm_se})
has the important property of satisfying what is known as the strong zero variance principle (SZVP),
which says that the solution of the eigenproblem will have no statistical uncertainty if the exact wave function
exists within the span of the current wave function and its first derivatives.
In practice this means that as an accurate wave function is approached, statistical uncertainty in the LM is greatly reduced.
This is a generalization of the standard VMC zero variance principle, in which the energy has no uncertainty if
the wave function ansatz itself is exact.
To see where the SZVP comes from, consider the following rearrangement of Eq.\ (\ref{eqn:lm_se}) in which the
matrices have been approximated stochastically as in the traditional LM (i.e. via Eq.\ (\ref{eqn:trad_mc_mat}))
\begin{align}
0 & =
\sum_{y=0}^{n_\mathrm{v}} 
(H_{xy} - E \hspace{0.8mm} S_{xy}) c_y
\label{eqn:lm_xy_pencil} \\
& \approx
\sum_{y=0}^{n_\mathrm{v}} 
\sum_{\bm{n}\in\xi}
\frac{\langle\Psi^x|\bm{n}\rangle} {\langle\Phi|\bm{n}\rangle}
\frac{\langle\bm{n}| (\hat{H}-E) |\Psi^y\rangle} {\langle\bm{n}|\Phi\rangle} c_y
\label{eqn:trad_szvp_middle} \\
& =
\sum_{\bm{n}\in\xi}
\frac{\langle\Psi^x|\bm{n}\rangle} {\langle\Phi|\bm{n}\rangle}
\frac{\langle\bm{n}| (\hat{H}-E) \sum_{y=0}^{n_\mathrm{v}} |\Psi^y\rangle c_y} {\langle\bm{n}|\Phi\rangle}
\label{eqn:trad_szvp}
\end{align}
If the exact wave function exists within the LM subspace, which is to say there is
a vector $\bm{c}$ such that
\begin{align}
(\hat{H}-E) \sum_{y=0}^{n_\mathrm{v}} |\Psi^y\rangle c_y = 0,
\label{eqn:exact_psi_in_fds}
\end{align}
then the terms in Eq.\ (\ref{eqn:trad_szvp}) vanish independently for every $\bm{n}$, and so
the exact energy $E$ and the vector $\bm{c}$ giving the exact wave function will be found during the diagonalization
of Eq.\ (\ref{eqn:lm_se}) regardless of which random sample $\xi$ is taken.
In other words, they will be found with zero variance.

Although the present approach does not satisfy the SZVP exactly, its deviation from the SZVP vanishes quadratically
as the exact wave function is approached.
To see why, replace $\bm{H}$ and $\bm{S}$ with Figure \ref{fig:matrix_tiles}'s stochastic approximations
and (without loss of generality) set $c_0$ = 1, at which point the deviation of Eq.\ (\ref{eqn:lm_xy_pencil}) from zero
(i.e.\ the deviation from the SZVP) becomes
\begin{align}
\eta_x & =
\sum_{\bm{n}\in\xi}
\frac{1} {|\langle\Phi|\bm{n}\rangle|^2}
\Bigg[
\langle\Psi^x|\bm{n}\rangle
\langle\bm{n}|(\hat{H}-E)|\Psi\rangle \hspace{1mm} +
\notag \\
& \quad \quad
\sum_{y=1}^{n_\mathrm{v}}
\langle\Psi^x|
\frac{\partial}{\partial \mu_y}
\Bigg(
e^{\hat{\mathcal{K}}}|\bm{n}\rangle\langle\bm{n}|e^{-\hat{\mathcal{K}}} (\hat{H}-E) |\Psi\rangle
\Bigg)
c_y \Bigg]_{\hat{\mathcal{K}}=0}
\notag \\
  & =
\sum_{\bm{n}\in\xi}
\frac{1} {|\langle\Phi|\bm{n}\rangle|^2}
\Bigg[
\langle\Psi^x|\bm{n}\rangle \langle\bm{n}| (\hat{H}-E) \sum_{y=0}^{n_\mathrm{v}} |\Psi^y\rangle c_y \hspace{1mm} +
\notag \\
& \quad \quad
\sum_{y=1}^{n_\mathrm{v}}
\langle\Psi^x|
\frac{\partial}{\partial \mu_y}
\Bigg(
e^{\hat{\mathcal{K}}}|\bm{n}\rangle\langle\bm{n}|e^{-\hat{\mathcal{K}}}
\Bigg)
(\hat{H}-E) |\Psi\rangle c_y \Bigg]_{\hat{\mathcal{K}}=0}
\notag
\end{align}
If we again assume that the (un-normalized) exact wave function
$|\Psi_0\rangle = |\Psi\rangle + \sum_{z=1}^{n_\mathrm{v}} |\Psi^z\rangle c_z$
exists in the LM subspace,
which implies that
\begin{align}
(\hat{H}-E) |\Psi\rangle
& = - \sum_{z=1}^{n_\mathrm{v}} (\hat{H}-E) |\Psi^z\rangle c_z,
\label{eqn:replace_exact_psi}
\end{align}
then the deviation from the SZVP
simplifies to
\begin{align}
\eta_x & =
- \sum_{y=1}^{n_\mathrm{v}} \sum_{z=1}^{n_\mathrm{v}} c_y c_z
\sum_{\bm{n}\in\xi}
\frac{Q^{(\bm{n})}_{xyz}} {|\langle\Phi|\bm{n}\rangle|^2}
\label{eqn:quad_szvp_dev} \\
Q^{(\bm{n})}_{xyz} & \equiv
\Bigg[
\langle\Psi^x|
\frac{\partial}{\partial \mu_y}
\Bigg(
e^{\hat{\mathcal{K}}}|\bm{n}\rangle\langle\bm{n}|e^{-\hat{\mathcal{K}}}
\Bigg)
(\hat{H}-E) |\Psi^z\rangle \Bigg]_{\hat{\mathcal{K}}=0}
\notag
\end{align}
Thus the deviation from the SZVP vanishes quadratically as
\text{$|$\reflectbox{$\bm{c}$}$|^2$}
with the difference 
\text{\reflectbox{$\bm{c}$}} between the current and exact wave functions.
This is in stark contrast to the previous quasi-Newton optimization strategy \cite{Neuscamman:2013:cjagp}
which lacked any kind of zero variance principle for the optimization updates, a fact that
likely explains our previous observation that greatly increased sample lengths compared to the
traditional LM were needed to stabilize the quasi-Newton approach.

Note that while it is possible to approximate CJAGP's $\bm{S}$ matrix at an $N^4$ per-sample cost
using the traditional LM's stochastic approach of Eq.\ (\ref{eqn:trad_mc_mat}),
doing so would violate even this quadratic approach to the SZVP when used together with Figure \ref{fig:matrix_tiles}'s
approximation for $\bm{H}$.
Indeed, we have observed that drastically larger sample sizes are required when one mixes the traditional method for
approximating $\bm{S}$ with our new method for approximating $\bm{H}$, and so we also approximate $\bm{S}$
via Figure \ref{fig:matrix_tiles} for the sake of reducing statistical uncertainty, even though this approximation
is more complicated.

\subsection{Avoiding Matrix Builds}
\label{sec:amb}

Although the LM typically works by first building the matrices $\bm{H}$ and $\bm{S}$ and then solving the
generalized eigenvalue problem of Eq.\ (\ref{eqn:lm_se}), Krylov subspace (KS) methods \cite{eigenvalue_templates_2000}
such as the Davidson \cite{Davidson:1975:davidson} or Arnoldi \cite{Arnoldi:1951:arnoldi} methods can be employed
to eschew the matrix builds altogether.
Such a KS approach has been used previously \cite{Neuscamman:2012:fast_sr} in the context of
stochastic reconfiguration \cite{Sorella:2001:SR,Sorella:2004:agp_sr},
and here we give some details for how such approaches can be generalized to both the traditional LM and
the newly proposed variant for CJAGP.
Instead of requiring the matrices to be built, KS methods typically only require the ability
to operate the matrix on an arbitrary vector, which in the context of either the LM or stochastic reconfiguration can
be advantageous when the number of wave function variables $n_\mathrm{v}$ becomes large.

In the traditional LM, a KS method will require evaluation of matrix-vector products $\bm{A} \bm{c}$ with the
stochastic matrix approximation given in Eq.\ (\ref{eqn:trad_mc_mat}):
\begin{align}
\sum_y A_{xy} c_y &\approx
\sum_{\bm{n}\in\xi} \frac{\langle\Psi^x|\bm{n}\rangle}{\langle\Psi|\bm{n}\rangle}
\sum_{y} \frac{\langle\bm{n}|\hat{A}|\Psi^y\rangle}{\langle\bm{n}|\Psi\rangle} c_y
\label{eqn:trad_lm_mv}
\end{align}
For wave functions like the JAGP \cite{Neuscamman:2013:jagp} for which the derivative vectors
$\langle\Psi^x|\bm{n}\rangle / \langle\Psi|\bm{n}\rangle$ and
$\langle\bm{n}|\hat{A}|\Psi^y\rangle / \langle\bm{n}|\Psi\rangle$ can be evaluated
efficiently, each sample's contribution to the overall matrix vector product
can be computed via a simple dot product.
If, for example, storing or communicating the matrix would be prohibitive, this
approach offers a lower-memory, lower-communication alternative.

For the approach proposed here for the CJAGP ansatz, the matrix vector product takes on two parts.
For the portion of the sum over $y$ covering the non-differentiated term, the pairing matrix derivatives,
and the Jastrow derivatives, the evaluation is the same as in Eq.\ (\ref{eqn:trad_lm_mv}).
For the portion of the sum in which $y$ runs over orbital rotation variables $K_{pq}$, we use
Eq.\ (\ref{eqn:new_Axy}) to write
\begin{align}
& \sum_{y\in\mathrm{orb.\ rot.}} A_{xy} c_y
\notag \\
& \hspace{6mm} \approx
\sum_{\bm{n}\in\xi} \frac{\langle\Psi^x|\bm{n}\rangle}{\langle\Phi|\bm{n}\rangle}
                                     \frac{\langle\bm{n}| \breve{A} |\Phi\rangle}{\langle\bm{n}|\Phi\rangle}
         +                           \frac{\langle\Psi^x| \breve{B} |\bm{n}\rangle}{\langle\Phi|\bm{n}\rangle}
                                     \frac{\langle\bm{n}| \hat{A} |\Phi\rangle}{\langle\bm{n}|\Phi\rangle}
\label{eqn:new_Axy_cy} \\
& \hspace{3mm} \breve{B} \equiv  \sum_{p<q} c_{pq} ( a^+_p a_q - a^+_q a_p )
\label{eqn:breve_B} \\
& \hspace{3mm} \breve{A} \equiv \left[ \hat{A}, \hspace{1mm} \breve{B} \right]
\label{eqn:breve_A}
\end{align}
In the definition of $\breve{B}$ we have relabeled the sum on $y$ over orbital rotations by the
orbital indices $p$ and $q$ that label the individual orbital rotation variables.
Crucially, because $\breve{B}$ is a one-electron operator, $\breve{A}$ is a two-electron operator
with exactly the same form as $\hat{A}$.
Moreover, the coefficients for $\breve{A}$ are independent of $\bm{n}$ and can thus be precomputed at an $N^5$ cost
\textit{before} the sample is taken, so that the actual per-sample cost of evaluating the first term in
Eq.\ (\ref{eqn:new_Axy_cy}) scales as only $N^4$.
The second term in Eq.\ (\ref{eqn:new_Axy_cy}) also has a per-sample cost scaling as $N^4$, as it amounts to
a weighted sum over the matrix elements of Eq.\ (\ref{eqn:new_Sxy}) scaled either by one (if $\bm{A} = \bm{S}$)
or by the JAGP local energy (if $\bm{A} = \bm{H}$).
Thus we see that in contrast to building the CJAGP LM matrices, which due to Eq.\ (\ref{eqn:expanded_commutator}) has
a per-sample cost scaling of $N^5$, operating these matrices on an arbitrary vector without building them has a
per-sample cost scaling of only $N^4$.
In large systems this reduced scaling could be an advantage, depending on how many matrix vector products are
required for the chosen Krylov subspace method.
Systems studied in this work are too small for this reduced scaling to be beneficial, but we present the
option of avoiding matrix builds anyways as it should be useful in future work.

\section{Results}
\label{sec:results}

\subsection{Computational Details}
\label{sec:comp_det}

CJAGP results were obtained using our own software for VMC in Hilbert space, with one- and two-electron
integrals for the Hamiltonian taken from Psi3 \cite{Psi3}.
Complete-active space self-consistent field (CASSCF) \cite{Werner:1985_1:mcscf,Werner:1985_2:mcscf},
full configuration interaction (FCI) \cite{Handy:1984:fci,Handy:1989:fci},
complete-active space second order perturbation theory (CASPT2) \cite{Werner:1996:caspt2},
and size-consistency-corrected multi-reference configuration interaction (MRCI+Q)
\cite{Knowles:1988:mrci,Werner:1988:mrci} results were obtained with Molpro \cite{MOLPRO_brief}.
Except for the (6e,12o) CASSCF result displayed in Figure \ref{fig:sco_631g_absolute_energy}, all other cases
of CASSCF, CASPT2, and MRCI+Q employed a minimal (6e,6o) active space containing the three pairs
of bonding/antibonding orbitals for the triple bonds of N$_2$ and [ScO]$^+$.
Results for restricted and unrestricted Hartree Fock (RHF and UHF) \cite{Szabo-Ostland} and coupled cluster with singles,
doubles, and perturbative triples (CCSD(T) and UCCSD(T))
\cite{BARTLETT:2007:cc_review} were obtained with QChem \cite{QChem:2006,QChem:2013}.
A 6-31G \cite{POPLE:1972:6-31g_basis} basis was used in all cases, and post-CASSCF methods (as well as CJAGP)
froze the N 1s, O 1s, and Sc 1s, 2s, and 2p orbitals.

In the optimization of our CJAGP wave function, some constraints were placed on the wave function to improve
the ease of convergence.
For the Jastrow factor, the coefficients were constrained to be symmetric between $\alpha$ and $\beta$ electrons,
so $J_{i_\alpha j_\alpha}=J_{i_\beta j_\beta}$ and $J_{i_\alpha j_\beta}=J_{i_\beta j_\alpha}$.
For the pairing matrix, we constrained $\bm{F}$ to be symmetric, resulting in an AGP reference with singlet spin.
Finally, we added further constraints to ensure Jastrow coefficients and pairing matrix elements that should be equal
by molecular symmetry were indeed equal.
For example, in N$_2$ and [ScO]$^+$ the Jastrow coefficients for equivalent s-p$_{\mathrm{x}}$ and s-p$_{\mathrm{y}}$ couplings
were constrained to be equal.

\begin{figure}[t]
\centering
\includegraphics[width=7.5cm,angle=270]{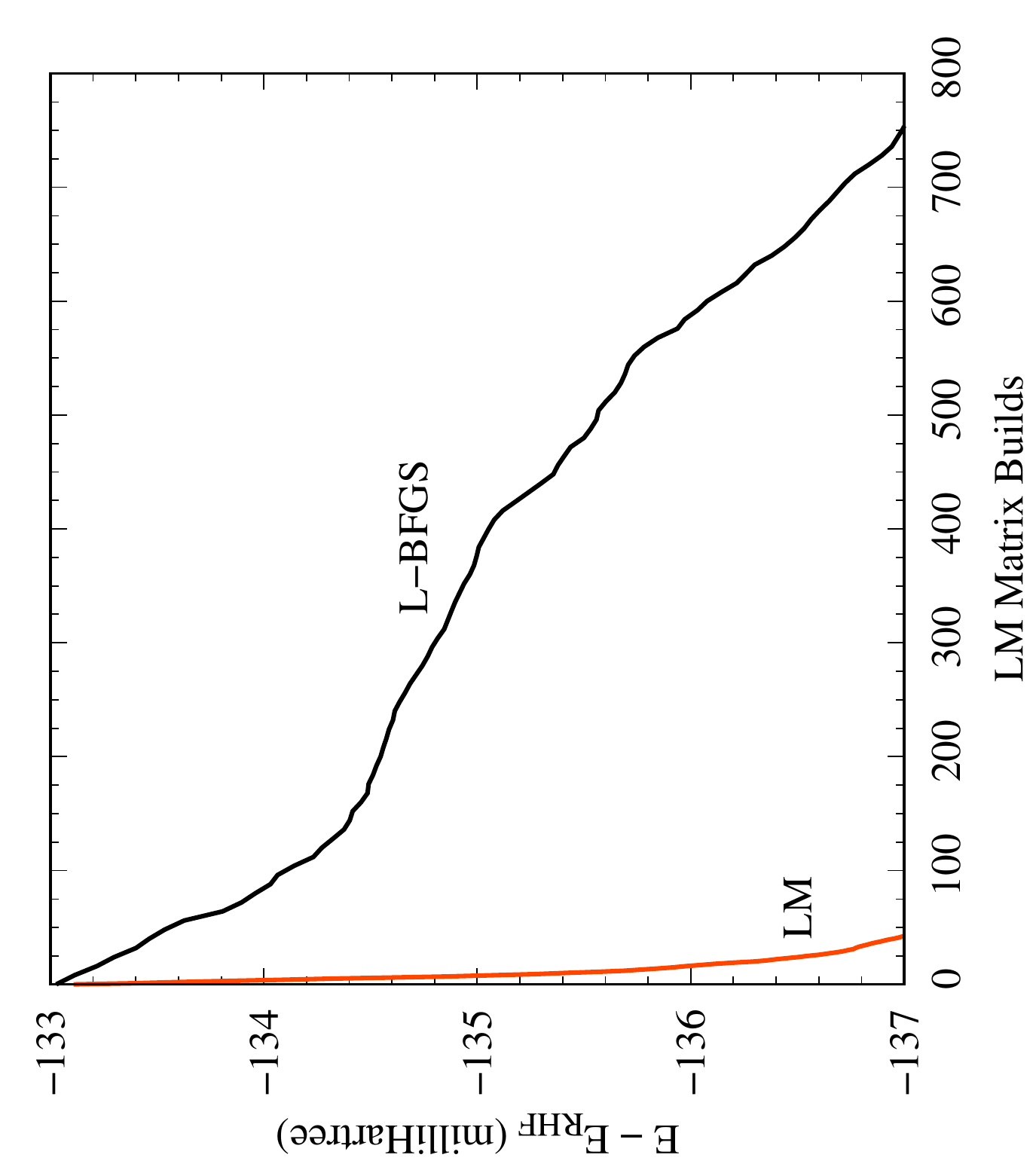}
\caption{Convergence of the last few mE$_{\mathrm{h}}$ of correlation energy for the LM and
         L-BFGS optimization approaches for H$_2$O in a 6-31G basis with $r_{\mathrm{OH}}=1.0\hspace{0.4mm}\text{\AA}$ and
         $\angle HOH=109.57^{\circ}$, plotted against the number of LM matrix builds completed after passing
         a correlation energy of -133 mE$_{\mathrm{h}}$.
         The converged CJAGP correlation energy is -137.3 mE$_{\mathrm{h}}$.
         See Section \ref{sec:convergence} for further details.
        }
\label{fig:h2o_631g_convergence}
\end{figure}

Note that the optimized CJAGP energy has two possible sources of statistical uncertainty.
First, there is the usual uncertainty when estimating the final wave function's energy using VMC.
Second, 
statistical uncertainty in the LM update direction $\text{\reflectbox{$\bm{c}$}}$ prevents
the optimal variable values from being found precisely.
In practice we observe the latter effect to be dominant, making the estimation of the overall method's
statistical uncertainty somewhat difficult, as we do not wish to run a large number of separate
optimizations at each molecular geometry to collect statistics.
Instead, we have fit CJAGP's energy error over the dissociation curves to a smooth third order
polynomial (e.g.\ see Figure \ref{fig:n2_631g_shifted_error}) and then estimated the statistical
uncertainty of the energies based on the deviations of the actual points from this smooth curve.
Assuming these deviations are normally distributed, we find 95\% confidence intervals of
$\pm 0.13$ kcal/mol in N$_2$ and $\pm 0.3$ kcal/mol in [ScO]$^+$.

\subsection{Convergence}
\label{sec:convergence}

\begin{figure}[t]
\centering
\includegraphics[width=7.5cm,angle=270]{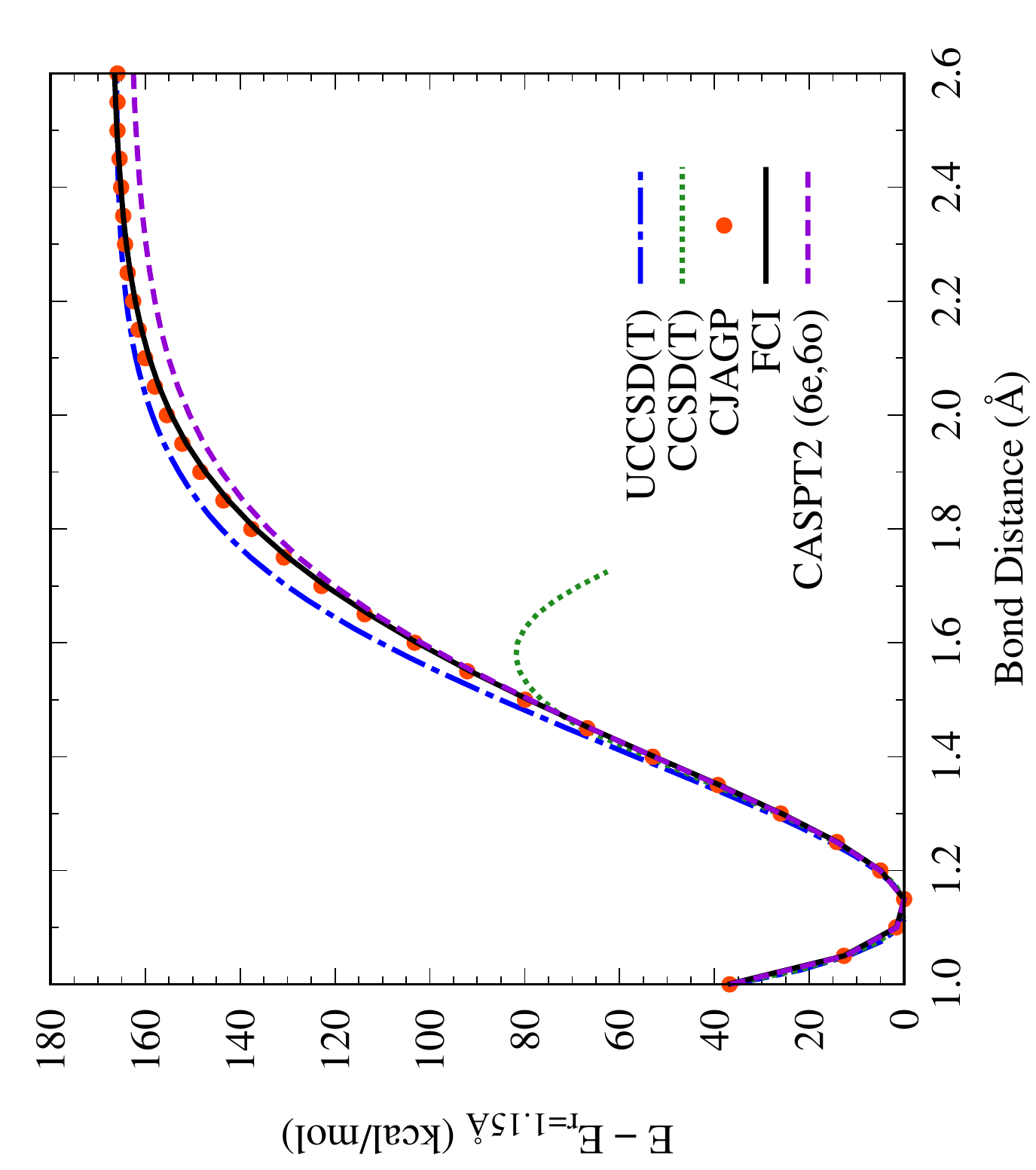}
\caption{Potential energy curves for N$_2$ dissociation in a 6-31G basis, with each curve shifted so that the
         zero of energy occurs at 1.15 \AA.  See Section \ref{sec:n2} for further details.
        }
\label{fig:n2_631g_shifted_energy}
\end{figure}

Figure \ref{fig:h2o_631g_convergence} shows, in H$_2$O near equilibrium, the convergence of the present LM approach compared to the previous \cite{Neuscamman:2013:cjagp} 
quasi-Newton L-BFGS optimization scheme.
The L-BFGS approach's idea was to optimize the orbital rotation $\hat{\mathcal{K}}$ on a surface $E(\bm{K})$ on which the other variables took on their optimal values
(i.e.\ the Jastrow and pairing variables for a given $\bm{K}$ were taken as those that minimized the energy for that $\bm{K}$).
In practice this surface was achieved by using the LM to reoptimize the Jastrow and pairing variables at each L-BFGS step,
and so we are able to compare the number of LM matrix builds required in that scheme to the number required by the present full LM approach.
While this comparison is somewhat imperfect as the previous LM matrix builds were less expensive than the present ones that also include the orbital rotation variables,
Figure \ref{fig:h2o_631g_convergence} nonetheless displays the stark contrast in optimization efficiency between the two approaches.

Note that this example was carried out under what might be called ``exact sampling'' (meaning that each configuration $\bm{n}$ was visited exactly once and its
contribution to averages scaled by the wave function weight $|\langle\bm{n}|\Psi\rangle|^2$) so that statistical uncertainty was not present.
In addition to being useful for debugging, such sampling allows us to test whether, independent of stochastic issues, the present LM outperforms L-BFGS,
and indeed it clearly does.
Note that the comparison becomes even more favorable for the present LM approach when a stochastic Markov-chain-based sample is used, as the LM
obeys a zero variance principle while the L-BFGS approach does not.
In practice, we therefore see that not only is the LM a superior optimization method, but that its inherently lower statistical uncertainty allows it to
operate effectively with much smaller sample sizes than are required to stabilize the previous L-BFGS approach.
One way to emphasize this advantage is to point out that in our previous study \cite{Neuscamman:2013:cjagp}, the stochastic sample sizes needed to stabilize the L-BFGS method were in all cases
larger than the Hilbert spaces themselves (note that this is not uncommon for stochastic approaches in small systems), whereas the present study's
sample lengths of 1.6$\times$10$^7$ for N$_2$ and 2.56$\times$10$^7$ for [ScO]$^+$ were in both cases smaller than the Hilbert spaces in question.

\begin{table*}[t]
\centering
\caption{Energies for the N$_2$ stretch in a 6-31G basis.
         FCI is reported in E$_{\mathrm{h}}$, with other methods reported as
         the difference from FCI in mE$_{\mathrm{h}}$.
         The last row gives the non-parallelity errors in mE$_{\mathrm{h}}$.
         See Section \ref{sec:n2} for further details.
        }
\begin{tabular*}{0.90\textwidth}{@{\extracolsep{\fill}} r r r r r r r r r }
\hline
\hline
  \multicolumn{1}{c}{$R$ (\AA)} &
  \multicolumn{1}{c}{FCI} &
  \multicolumn{1}{c}{RHF} &
  \multicolumn{1}{c}{UHF} &
  \multicolumn{1}{c}{CCSD(T)} &
  \multicolumn{1}{c}{UCCSD(T)} &
  \multicolumn{1}{c}{CASSCF} &
  \multicolumn{1}{c}{CASPT2} &
  \multicolumn{1}{c}{CJAGP} \\
\hline
           1.00 &      -109.0467 &          211.4 &          211.4 &           -0.8 &           -0.8 &           85.5 &           13.7 &            6.8 \\
           1.05 &      -109.0857 &          223.3 &          223.3 &           -0.5 &           -0.5 &           86.5 &           14.0 &            7.4 \\
           1.10 &      -109.1034 &          235.7 &          235.7 &           -0.2 &           -0.2 &           87.4 &           14.3 &            7.5 \\
           1.15 &      -109.1059 &          248.8 &          247.5 &            0.2 &            0.9 &           88.3 &           14.5 &            7.4 \\
           1.20 &      -109.0981 &          262.4 &          252.9 &            0.6 &            2.3 &           89.2 &           14.7 &            7.6 \\
           1.25 &      -109.0835 &          276.6 &          252.9 &            1.1 &            3.5 &           90.0 &           14.8 &            7.6 \\
           1.30 &      -109.0648 &          291.3 &          249.1 &            1.5 &            4.6 &           90.8 &           14.8 &            7.8 \\
           1.35 &      -109.0438 &          306.6 &          242.7 &            2.1 &            5.8 &           91.6 &           14.8 &            7.9 \\
           1.40 &      -109.0221 &          322.4 &          234.6 &            2.6 &            7.2 &           92.3 &           14.6 &            8.1 \\
           1.45 &      -109.0005 &          338.8 &          225.3 &            3.0 &            8.7 &           93.0 &           14.4 &            8.5 \\
           1.50 &      -108.9797 &          352.9 &          215.1 &           -4.2 &           10.3 &           93.5 &           14.1 &            8.6 \\
           1.55 &      -108.9602 &          363.1 &          203.9 &          -16.3 &           11.9 &           93.9 &           13.6 &            8.5 \\
           1.60 &      -108.9423 &          370.5 &          191.9 &          -33.6 &           13.2 &           94.0 &           13.0 &            8.3 \\
           1.65 &      -108.9260 &          376.2 &          179.1 &          -56.4 &           13.9 &           94.0 &           12.4 &            8.8 \\
           1.70 &      -108.9116 &          380.9 &          166.2 &          -84.9 &           14.0 &           93.6 &           11.6 &            8.9 \\
           1.75 &      -108.8989 &          385.2 &          153.5 &                &           13.5 &           92.9 &           10.9 &            8.8 \\
           1.80 &      -108.8880 &          389.6 &          141.5 &                &           12.4 &           91.9 &           10.2 &            8.8 \\
           1.85 &      -108.8788 &          394.3 &          130.6 &                &           11.1 &           90.7 &            9.5 &            9.0 \\
           1.90 &      -108.8711 &          399.5 &          120.9 &                &            9.8 &           89.3 &            9.0 &            9.1 \\
           1.95 &      -108.8648 &          405.2 &          112.5 &                &            8.5 &           87.7 &            8.6 &            8.8 \\
           2.00 &      -108.8597 &          411.4 &          105.2 &                &            7.3 &           86.2 &            8.3 &            8.9 \\
           2.05 &      -108.8556 &          418.0 &           99.1 &                &            6.3 &           84.8 &            8.1 &            8.9 \\
           2.10 &      -108.8523 &          424.9 &           93.9 &                &            5.4 &           83.4 &            8.0 &            8.8 \\
           2.15 &      -108.8496 &          432.1 &           89.6 &                &            4.6 &           82.1 &            8.0 &            8.3 \\
           2.20 &      -108.8476 &          439.4 &           86.1 &                &            3.9 &           81.0 &            7.9 &            8.2 \\
           2.25 &      -108.8459 &          446.7 &           83.1 &                &            3.2 &           80.0 &            7.9 &            8.3 \\
           2.30 &      -108.8446 &          454.1 &           80.6 &                &            2.6 &           79.2 &            8.0 &            7.9 \\
           2.35 &      -108.8435 &          461.5 &           78.6 &                &            2.1 &           78.5 &            8.0 &            7.4 \\
           2.40 &      -108.8427 &          468.7 &           76.9 &                &            1.6 &           77.8 &            8.0 &            7.3 \\
           2.45 &      -108.8420 &          475.8 &           75.5 &                &            1.2 &           77.3 &            8.1 &            7.1 \\
           2.50 &      -108.8414 &          482.7 &           74.3 &                &            0.8 &           76.9 &            8.1 &            7.2 \\
           2.55 &      -108.8410 &          489.5 &           73.4 &                &            0.5 &           76.5 &            8.1 &            6.8 \\
           2.60 &      -108.8406 &          496.1 &           72.5 &                &            0.2 &           76.1 &            8.2 &            6.4 \\
\hline
            NPE &            N/A &          284.6 &          180.3 &           87.9 &           14.8 &           17.9 &            6.9 &            2.7 \\
\hline
\hline
\end{tabular*}
\label{tab:n2_energies}
\end{table*}

\subsection{N$_2$}
\label{sec:n2}

\begin{figure}[t]
\centering
\includegraphics[width=7.5cm,angle=270]{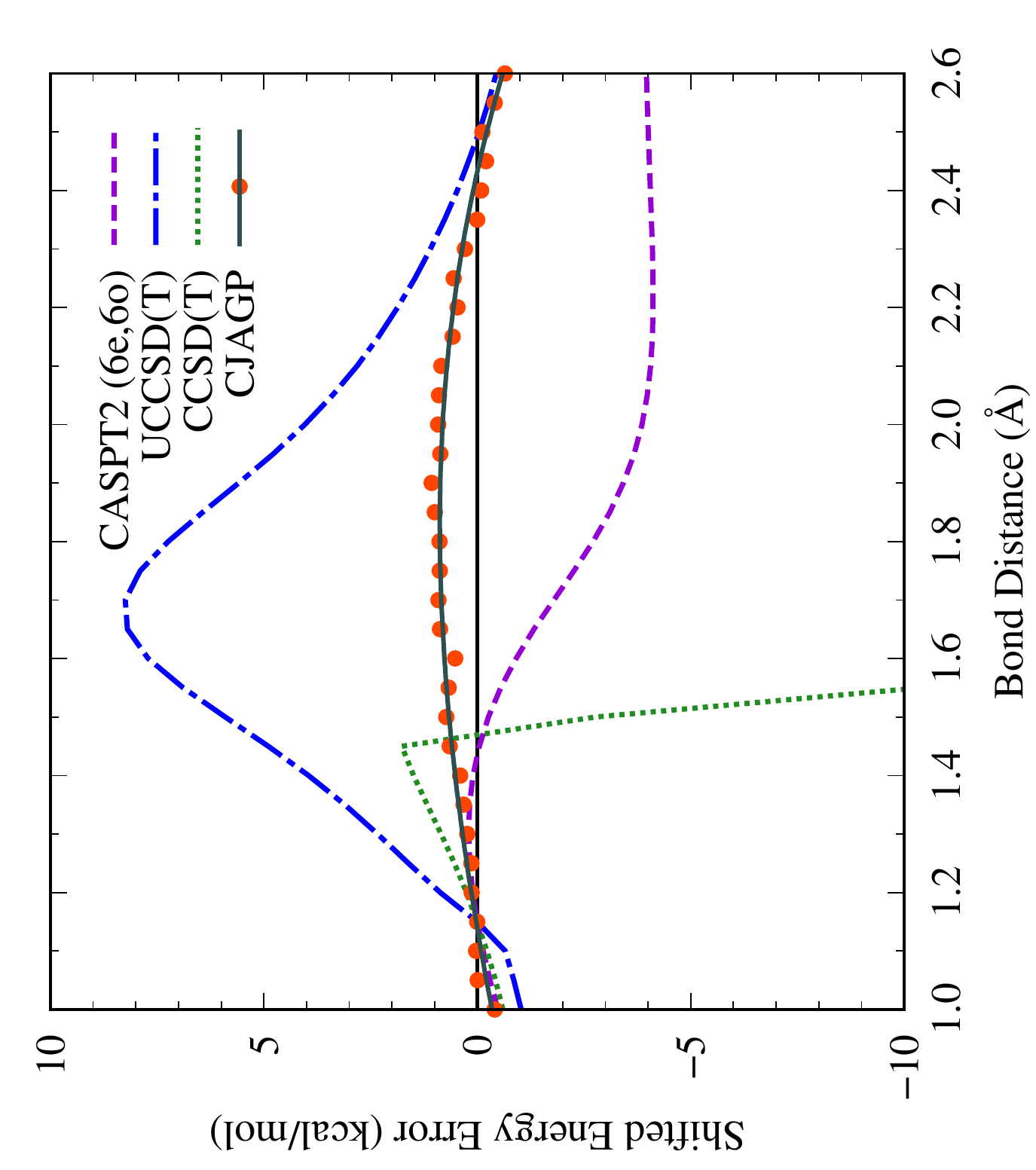}
\caption{Energy deviations from FCI during N$_2$ dissociation in a 6-31G basis, with each curve shifted by a constant
         so that it crosses zero at a bond distance of 1.15 \AA.
         For CJAGP, the line is a cubic polynomial fit to the points to give a sense of statistical uncertainty.
         See Section \ref{sec:n2} for further details.
        }
\label{fig:n2_631g_shifted_error}
\end{figure}

The dissociation of the nitrogen dimer's triple bond has long been used as a benchmark for multi-reference
methods in quantum chemistry.
As was seen previously \cite{Neuscamman:2013:cjagp} in H$_2$O and HF, the limited CC-like nature of the CJAGP's orbital-rotated Jastrow
factor appears to capture a large fraction of the dynamic correlation energy while maintaining the ability to help capture static
correlation in conjunction with the geminal power \cite{Neuscamman:2015:subtractive_jagp}.
As we see in the N$_2$ results (Figures \ref{fig:n2_631g_shifted_energy} and \ref{fig:n2_631g_shifted_error} and Table \ref{tab:n2_energies})
these features allow CJAGP to vastly outperform single-reference methods like CCSD(T) and UCCSD(T).
Here the catastrophic failure of CCSD(T) may be attributed both to the poor quality of its RHF reference
(whose instabilities towards spatial
symmetry breaking are responsible for the kink in its potential curve) and to the tendency of
spurious interactions between its singlet and triplet amplitude channels to overcorrelate in the strongly correlated regime \cite{Scuseria:2015:ccd0}.
Note that the issue of spatial symmetry breaking in the RHF might be avoided by enforcing spatial symmetry throughout the dissociation, but for N$_2$ we have
chosen to present the CCSD(T) results for the minimum energy RHF reference as found via stability analyses \cite{Pople:1977:hf_stability}.
In contrast, CJAGP avoids these issues thanks to its more flexible reference function and the variational nature of its evaluation, which guarantees
that spurious couplings between its cluster amplitudes cannot lead to an overcorrelation catastrophe.

\begin{table*}[t]
\centering
\caption{Energies for the [ScO]$^+$ stretch in a 6-31G basis.
         MRCI+Q is reported in E$_{\mathrm{h}}$, with other methods reported as
         the difference from MRCI+Q in mE$_{\mathrm{h}}$.
         The last row gives the non-parallelity errors in mE$_{\mathrm{h}}$.
         See Section \ref{sec:sco} for further details.
        }
\begin{tabular*}{0.90\textwidth}{@{\extracolsep{\fill}} r r r r r r r r r }
\hline
\hline
  \multicolumn{1}{c}{$R$ (\AA)} &
  \multicolumn{1}{c}{MRCI+Q} &
  \multicolumn{1}{c}{RHF} &
  \multicolumn{1}{c}{UHF} &
  \multicolumn{1}{c}{CCSD(T)} &
  \multicolumn{1}{c}{UCCSD(T)} &
  \multicolumn{1}{c}{CASSCF (6e,6o)} &
  \multicolumn{1}{c}{CASPT2} &
  \multicolumn{1}{c}{CJAGP} \\
\hline
            1.5 &      -834.6354 &          345.5 &          345.5 &            0.8 &            0.8 &          202.9 &           25.5 &           69.4  \\
            1.6 &      -834.6631 &          356.3 &          356.3 &            0.6 &            0.6 &          206.0 &           26.0 &           69.6  \\
            1.7 &      -834.6688 &          367.5 &          367.5 &            0.4 &            0.4 &          209.3 &           26.6 &           69.0  \\
            1.8 &      -834.6607 &          379.6 &          357.9 &            0.1 &            8.9 &          212.8 &           27.1 &           68.3  \\
            1.9 &      -834.6445 &          392.9 &          340.5 &           -0.4 &            7.6 &          216.4 &           27.4 &           68.7  \\
            2.0 &      -834.6243 &          407.0 &          324.2 &           -0.9 &            5.3 &          219.9 &           27.5 &           69.5  \\
            2.1 &      -834.6024 &          420.6 &          310.1 &            4.7 &            5.3 &          223.1 &           27.1 &           68.5  \\
            2.2 &      -834.5806 &          427.4 &          298.7 &           14.4 &            8.7 &          225.7 &           26.1 &           69.3  \\
            2.3 &      -834.5600 &          429.1 &          289.8 &           14.2 &           12.5 &          227.3 &           24.2 &           70.5  \\
            2.4 &      -834.5413 &          427.9 &          281.8 &            7.6 &           15.2 &          228.3 &           21.6 &           69.7  \\
            2.5 &      -834.5248 &          425.2 &          266.5 &           -6.3 &           14.2 &          228.7 &           17.0 &           70.4  \\
            2.6 &      -834.5106 &          421.6 &          252.7 &          -34.4 &           12.0 &          228.2 &                &           71.6  \\
\hline
            NPE &            N/A &           83.5 &          114.7 &           48.8 &           14.9 &           25.7 &           10.5 &            3.3 \\
\hline
\hline
\end{tabular*}
\label{tab:sco_energies}
\end{table*}

More significantly, CJAGP outperforms CASPT2, one of the most affordable
and most commonly used multi-reference methods in quantum chemistry.
Both its absolute and relative energies show improvements compared to those of CASPT2, with the relative energies being particularly
accurate:  the non-parallelity error (NPE, the difference between the highest and lowest deviations) relative to FCI is less than 2 kcal/mol
and less than half that of CASPT2.
These improvements are especially significant when one considers that CASPT2's cost scales exponentially due to
its complete active space reference, while CJAGP's cost scales only polynomially.

In light of the Jastrow factor's CC-like form and the geminal power's multi-reference nature, it is interesting to compare CJAGP to the
performance one might expect from the ideal of a variational singles-and-doubles CC method based on a complete active space self consistent field
(CASSCF) wave function reference.
As such a theory should outperform even MRCI+Q,
one would expect absolute accuracies to be within 1 or 2 mE$_{\mathrm{h}}$ of FCI (see e.g.\ \cite{Neuscamman:2009:qct}).
Unsurprisingly, given that both its cluster operator and its AGP reference function are more constrained than this ideal,
CJAGP does not achieve such accuracies in the absolute energy.
Its relative energies are nonetheless quite accurate, suggesting that the missing details that would account for the last few percent of
the correlation energy are being left out consistently at all geometries.
If supplied with a trial function as accurate as CJAGP, diffusion Monte Carlo \cite{FouMitNeeRaj-RMP-01} would be well placed to capture
these final details.
One very interesting question going forward is thus whether a real-space Jastrow factor can be devised to replicate the CC qualities of
the orbital-rotated Hilbert-space Jastrow.

\subsection{ScO Cation}
\label{sec:sco}

\begin{figure}[b]
\centering
\includegraphics[width=7.5cm,angle=270]{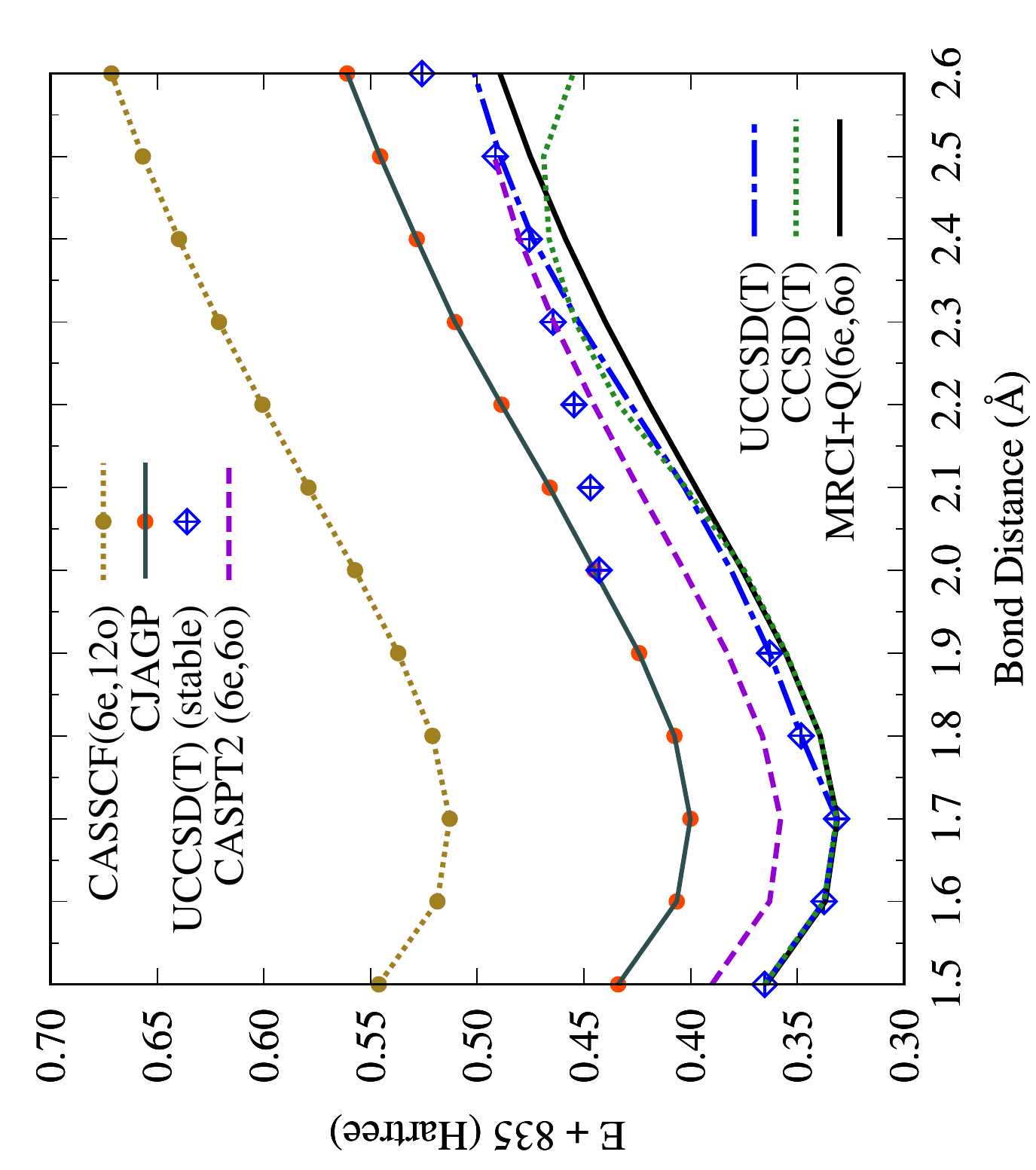}
\caption{Total energies during the dissociation of [ScO]$^+$ in a 6-31G basis.
         See Section \ref{sec:sco} for further details.
        }
\label{fig:sco_631g_absolute_energy}
\end{figure}

\begin{figure}[t]
\centering
\includegraphics[width=7.5cm,angle=270]{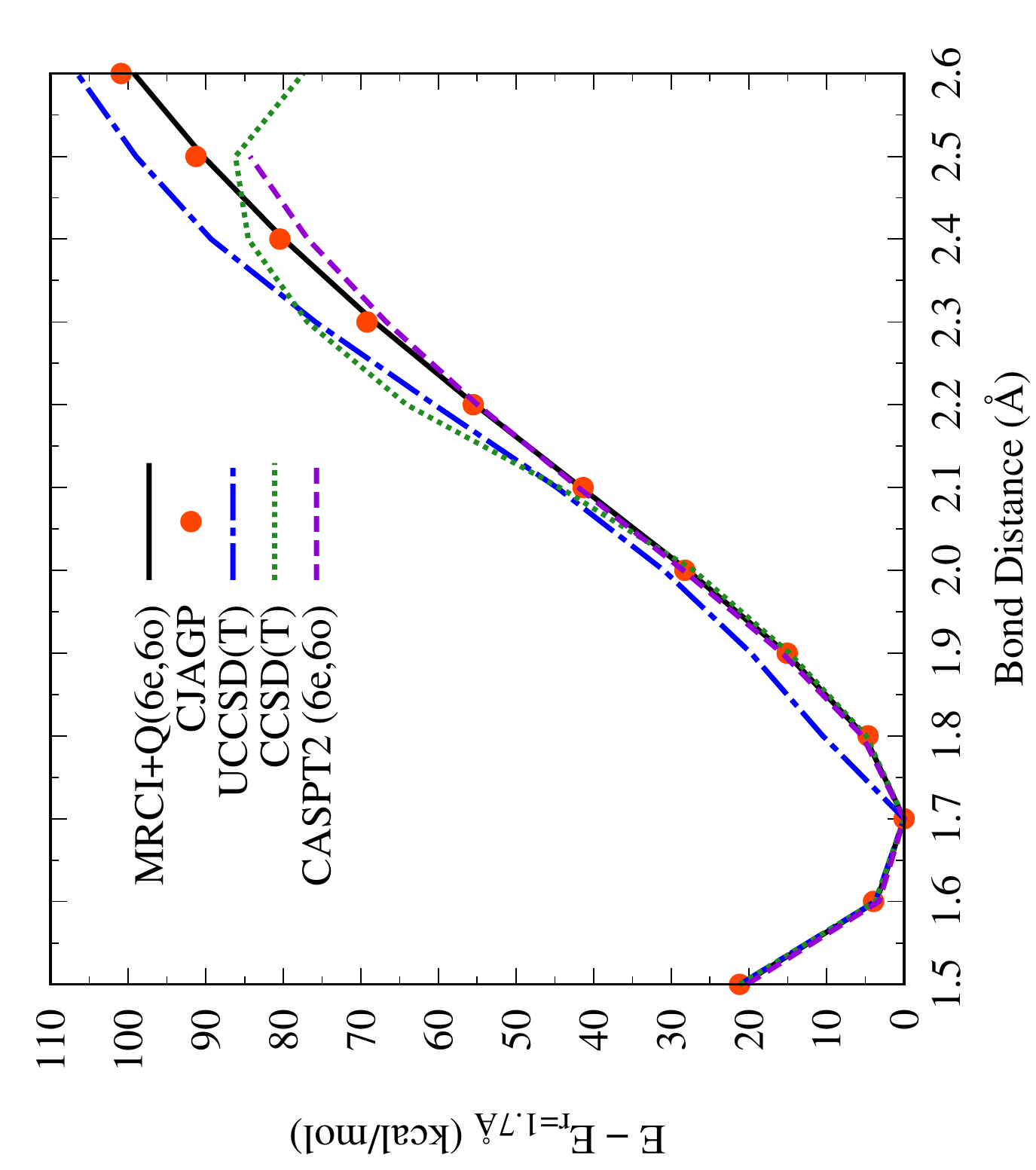}
\caption{Energy curves during the dissociation of [ScO]$^+$ in a 6-31G basis,
         with each curve shifted so that the zero of energy occurs at 1.7 \AA.
         See Section \ref{sec:sco} for further details.
        }
\label{fig:sco_631g_shifted_energy}
\end{figure}

\begin{figure}[t]
\centering
\includegraphics[width=7.5cm,angle=270]{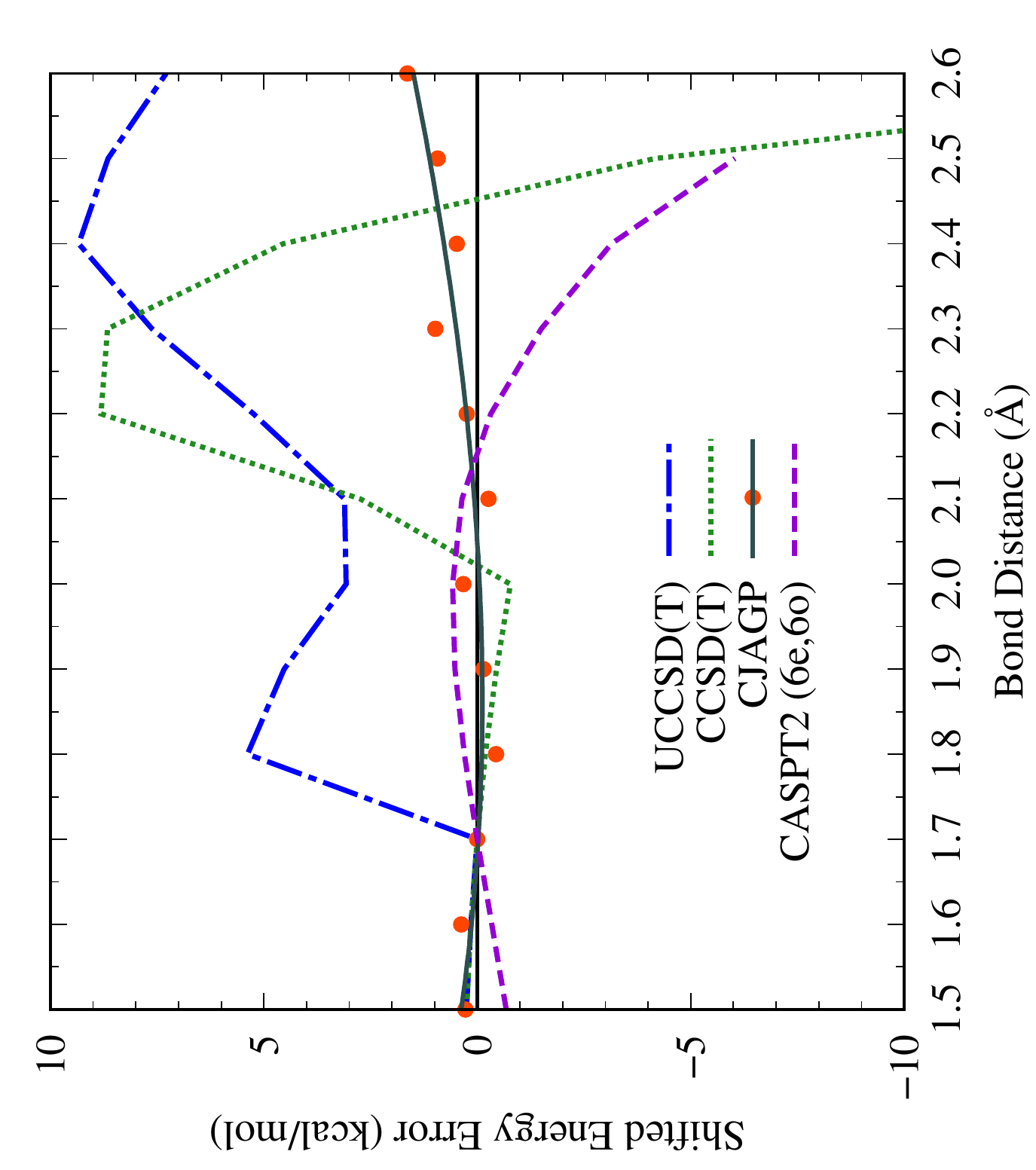}
\caption{Energy deviations from MRCI+Q (6e,6o) during [ScO]$^+$ dissociation in a 6-31G basis, with each curve shifted by a constant
         so that it crosses zero at a bond distance of 1.7 \AA.
         For CJAGP, the line is a cubic polynomial fit to the points to give a sense of statistical uncertainty.
         See Section \ref{sec:sco} for further details.
        }
\label{fig:sco_631g_shifted_error}
\end{figure}

Due to the importance of transition metals in catalysis and materials science, and the tendency of metal-oxygen bonds to exhibit
strong electron correlations, theoretical approaches that can deal successfully with such correlations are a high priority.
As an initial foray into this regime, we have tested CJAGP on the triple-bond dissociation of [ScO]$^+$.
At first glance, this cation appears quite similar to N$_2$ in that it also contains one $\sigma$ and two $\pi$ bonds.
In practice, however, its dissociation is even more fraught, with UCCSD(T) becoming qualitatively unreliable and minimal-active-space
CASPT2 exhibiting intruder state problems.
As seen in Figure \ref{fig:sco_631g_absolute_energy}, CCSD(T) exhibits its typical failure during multiple bond stretching.
UCCSD(T) fares little better, being beset by a Coulson-Fischer point cusp near equilibrium where RHF and UHF separate
as well as multiple low-lying UHF determinants as the bond is stretched.
If one uses stability analyses to ensure that UCCSD(T) is always based on the lowest energy UHF solution, the result is
a UCCSD(T) curve (UCCSD(T) (stable)) with multiple discontinuities.
These discontinuities can be avoided by always using the UHF solution with character most similar to the $R=1.9$ \AA\ UHF state,
as we have done for the data labeled UCCSD(T) in Figures \ref{fig:sco_631g_absolute_energy}-\ref{fig:sco_631g_shifted_error}
and Table \ref{tab:sco_energies}, but even in this case UCCSD(T) displays a NPE of 9.3 kcal/mol.
One should bear in mind that without benchmark results it would be difficult to know whether this UHF determinant or
the lower energy determinants found through stability analyses were the more reasonable starting points, and so it is hard to
recommend the use of UCCSD(T) for predicting energy profiles when stretching transition-metal-oxide bonds.

When based on the triple-bond's minimal (6e,6o) active space, CASPT2 proves more reliable than CC and achieves a smaller 6.6 kcal/mol NPE.
However, this CASPT2 approach failed to converge at $R=2.6$ \AA\ due to the presence of an intruder state.
One could overcome this problem with either a larger-than-minimal active space or through the use of level shifts \cite{Andersson:1995:caspt2_level_shift},
but the former may become untenable in larger transition metal systems while the latter introduces an uncontrolled free parameter.

As in N$_2$, the active-space-free CJAGP improves on the relative energy of CASPT2 with a NPE of just 2.1 kcal/mol, as seen in
Figures \ref{fig:sco_631g_shifted_energy} and \ref{fig:sco_631g_shifted_error}.
However, as seen in Figure \ref{fig:sco_631g_absolute_energy} and Table \ref{tab:sco_energies}, the absolute energy errors for CJAGP
are now much larger than they were in N$_2$.
While Figure \ref{fig:sco_631g_absolute_energy} reveals that CJAGP recovers significantly more correlation energy than even a
full valence (6e,12o) CASSCF approach, it is still missing roughly 70 mE$_{\mathrm{h}}$ relative to the benchmark MRCI+Q.
As discussed in Section \ref{sec:n2}, this performance is inferior to what one would expect from a (currently non-existent)
CASSCF-based variational CC.
Given the excellent shape of CJAGP's potential energy curve (again, NPE is only 2.1 kcal/mol), we do not think the issue lies
with the multi-reference Jastrow-AGP combination but instead suspect the missing correlation
energy is due to the limited flexibility of the Jastrow operator's CC form (Eq.\ (\ref{eqn:cluster_amp})) when compared to a full CC doubles operator.
In other words, we suspect that the limited CC flexibility leads to a limited dynamic correlation recovery, although one that
is surprising well balanced across different geometries.
As for N$_2$, these results strongly suggest that excellent accuracies could be achieved if DMC could use a trial function of
CJAGP quality, as DMC is excellent at recovering dynamic correlation details when supplied with a qualitatively correct trial function \cite{TouUmr-JCP-08}.
Indeed, based on its energy results, CJAGP should be an even better DMC trial function than full-valence CASSCF,
and so we feel further motivated to investigate this exciting possibility.

As a final note, we would like to point out that beyond 2.6\AA, the CJAGP optimization failed to converge to a good singlet,
likely because at around this geometry a singlet-triplet crossing occurs and the singlet is no longer the ground state
\cite{Mavridis:2010:tmo_ScO_TiO_CrO_MnO}.
While we hope to investigate CJAGP's prospects for the direct, variational targeting of
excited states \cite{Zhao:2016:evp_qmc} in the future, we have limited ourselves here to bond distances below 2.6\AA\ for which
the singlet is the ground state.

\section{Conclusions}
\label{sec:conclusions}

We have presented an improved LM optimization scheme for the CJAGP ansatz that achieves an $N^5$
per-sample cost scaling that drops to $N^4$ if Krylov subspace methods are employed.
This LM optimization obeys the strong zero variance principle in a quadratic sense, and is thus
vastly more statistically efficient than the previously employed quasi-Newton approach.
In practice this improved optimization scheme has led to drastic reductions in the sample sizes
and optimization steps required for variational energy minimization.
The key theoretical development facilitating these improvements was the use of an alternative
stochastic resolution of the identity in the estimation of the LM matrices or matrix vector
products.

With this improved optimization scheme, we showed that CJAGP is vastly more reliable than
traditional single-reference CC in two challenging triple-bond dissociations, one involving
a transition metal.
Further, we showed that for relative energies, the polynomial-cost, active-space-free CJAGP also outperformed the
exponentially scaling, active-space-based CASPT2 method.
In both examples, the CJAGP relative energies were substantially more accurate than its absolute energy,
suggesting to us that the limited flexibility of its cluster operator ($O(N^2)$ variables vs the traditional $O(N^4$))
prevented the capture of the finer details of dynamic correlation in a way that was well balanced across
different geometries.

Our findings in this study suggest two important avenues for future investigation.
First, given that CJAGP appears to be a better trial function starting point than even a full-valence CASSCF reference,
it would be highly desirable to combine it with diffusion Monte Carlo.
This is not entirely trivial given that currently the CJ operator exists only in Hilbert (rather than real) space,
but we look forward to investigating how its success may inform real space ansatz development.
Second, our practical experience in applying CJAGP is making it increasingly clear that, in Hilbert space, the
primary issue that will constrain
the use of the CJAGP in the future is the fact that in its current form it must at each sample loop over a large slice
of the two-electron integrals.
As there has been much success in simplifying the handling of two-electron integrals in other areas of quantum
chemistry, either by tensor decomposition or by screening, we look forward to the possibility of similar efficiency gains 
in the context of the CJAGP.

\section{Acknowledgments}
\label{sec:acknowledgments}

Part of this work was performed under the auspices of the U.S. Department of Energy by Lawrence Livermore National Laboratory
under Contract DE-AC52-07NA27344.
We also acknowledged support from the University of California.

\appendix

\section{Evaluating Eq.\ (\ref{eqn:new_Sxy})}
\label{sec:oroc}

Here we give details on how Eq.\ (\ref{eqn:new_Sxy}) can be evaluated efficiently,
assuming that $p$ and $q$ correspond to $\alpha$ spin-orbitals.
The $\beta$ spin case follows exactly the same logic.
First, consider the case where $\mu_x$ is an orbital rotation variable, $\mu_x=K_{rs}$.
For each sampled configuration, this case requires evaluation of terms of the form
\begin{align}
\frac{\langle\Psi^x|a^+_p a_q|\bm{n}\rangle}{\langle\Phi|\bm{n}\rangle}
&= 
\frac{\langle\Phi| (a^+_s a_r - a^+_r a_s) a^+_p a_q|\bm{n}\rangle}{\langle\Phi|\bm{n}\rangle}
\label{eqn:pqrs_ratio}
\end{align}
which for the different values of $p,q,r,s$ amount to $O(N^4)$ double excitation ratios.
Like those of Eq.\ (\ref{eqn:double_excite_ratios}), these may be evaluated
for a total per-sample cost that scales as $N^4$.
Note that for terms with $r=p$ or $s=p$, the anti-commutation rules involved in rearranging the
creation and destruction operators to match Eq.\ (\ref{eqn:double_excite_ratios}) 
will also generate single excitation ratios, but these do not change the cost scaling
(see Eq.\ (\ref{eqn:pq_ratio}) below).

It remains to consider the case where $\mu_x$ is a pairing matrix element or Jastrow coefficient.
In this case it is helpful to rewrite the required term as
\begin{align}
\frac{\langle\Psi^x|a^+_p a_q|\bm{n}\rangle}{\langle\Phi|\bm{n}\rangle}
&=
\frac{\partial}{\partial \mu_x}
\left(
\frac{\langle\Phi|a^+_p a_q|\bm{n}\rangle}{\langle\Phi|\bm{n}\rangle}
\right)
\notag \\
& \quad \quad + \frac{\langle\Phi|a^+_p a_q|\bm{n}\rangle}{\langle\Phi|\bm{n}\rangle}
  \frac{\langle\Phi^x|\bm{n}\rangle}{\langle\Phi|\bm{n}\rangle}.
\label{eqn:grrrrarrrr}
\end{align}
From Eqs.\ (28-31) of Ref.\ \cite{Neuscamman:2013:jagp}, one can see that JAGP
single excitation ratios may be evaluated as
\begin{align}
\frac{\langle\Phi|a^+_p a_q|\bm{n}\rangle}{\langle\Phi|\bm{n}\rangle}
= (\bm{R}\bm{\Theta})_{pq} \exp( K^\alpha_p - K^\alpha_q - J^{\alpha\alpha}_{qp} ),
\label{eqn:pq_ratio}
\end{align}
where $\bm{R}$ is the unoccupied-occupied block of the pairing matrix,
$\bm{\Theta}$ is the inverse of the occupied-occupied block of the pairing matrix,
and $K^\alpha_p$ and $K^\alpha_q$ are Ref.\ \cite{Neuscamman:2013:jagp}'s Jastrow intermediates
(each of which is a simple sum over Jastrow factor coefficients).
As $\bm{\Theta}$ and the product $\bm{R}\bm{\Theta}$ are already evaluated for the JAGP LM and are
thus readily available, the ratios in Eq.\ (\ref{eqn:pq_ratio}) may all be evaluated for an
additional per-sample cost scaling as $N^2$.
These ratios in hand, and recognizing that the pairing matrix and Jastrow derivative ratios
$\langle\Phi^x|\bm{n}\rangle/\langle\Phi|\bm{n}\rangle$ are also already available,
we see that the last term in Eq.\ (\ref{eqn:grrrrarrrr}) may be evaluated for a per-sample cost
scaling as $N^4$.
All that remains now is the first term on the right hand side of Eq.\ (\ref{eqn:grrrrarrrr}),
which requires derivatives of Eq.\ (\ref{eqn:pq_ratio}) with respect to pairing matrix
and Jastrow variables.
In the Jastrow variable case, these are quite trivial, working out to
$\pm \langle\Phi|a^+_p a_q|\bm{n}\rangle / \langle\Phi|\bm{n}\rangle$
if the Jastrow variable appears in the exponential
(remember the intermediates are just sums of Jastrow variables)
and zero if it does not.
For pairing matrix elements that are part of the occupied-unoccupied block $\bm{R}$ for the current
configuration $\bm{n}$, these derivatives are
\begin{align}
& \frac{\partial}{\partial R_{ai}}
\left(
\frac{\langle\Phi|a^+_p a_q|\bm{n}\rangle}{\langle\Phi|\bm{n}\rangle}
\right)
\notag \\
& \qquad = \delta_{ap} \Theta_{iq} \exp( K^\alpha_p - K^\alpha_q - J^{\alpha\alpha}_{qp} ).
\label{eqn:occ_unocc_ders}
\end{align}
For pairing matrix elements that are part of the occupied-occupied block $\bm{\mathcal{O}}$
for which $\bm{\Theta}$ is the matrix inverse, these derivatives are
\begin{align}
& \frac{\partial}{\partial \mathcal{O}_{ij}}
\left(
\frac{\langle\Phi|a^+_p a_q|\bm{n}\rangle}{\langle\Phi|\bm{n}\rangle}
\right)
\notag \\
& \qquad = -(\bm{R}\bm{\Theta})_{pi} \Theta_{jq} \exp( K^\alpha_p - K^\alpha_q - J^{\alpha\alpha}_{qp} ).
\label{eqn:occ_occ_ders}
\end{align}
For other pairing matrix elements, on which the single excitation ratios do not depend, these
derivatives are zero.
In conclusion, whether considering orbital rotation variables via Eq.\ (\ref{eqn:pqrs_ratio}) or pairing matrix or
Jastrow factor variables via Eq.\ (\ref{eqn:grrrrarrrr}), all of the components for a sampled configuration's 
contribution to $\bm{S}$ via Eq.\ (\ref{eqn:new_Sxy}) may be evaluated at a cost that scales as $N^4$.

\bibliographystyle{aip}
\bibliography{cjagp_lm.bib}

\begin{thebibliography}{10}

\bibitem{Chan:2011:dmrg_in_chem}
G.~K.-L. Chan and S.~Sharma,
\newblock Annu. Rev. Phys. Chem. {\bf 62}, 465 (2011).

\bibitem{Booth:2015:mcscf_fciqmc}
R.~E. Thomas, Q.~Sun, A.~Alavi, and G.~H. Booth,
\newblock J. Chem. Theory Comput. {\bf 11}, 5316 (2015).

\bibitem{Neuscamman:2013:cjagp}
E.~Neuscamman,
\newblock J. Chem. Phys. {\bf 139}, 181101 (2013).

\bibitem{BARTLETT:2007:cc_review}
R.~J. Bartlett and M.~Musia{\l},
\newblock Rev. Mod. Phys. {\bf 79}, 291 (2007).

\bibitem{Neuscamman:2013:jagp}
E.~Neuscamman,
\newblock J. Chem. Phys. {\bf 139}, 194105 (2013).

\bibitem{Nightingale:2001:linear_method}
M.~P. Nightingale and V.~Melik-Alaverdian,
\newblock Phys. Rev. Lett. {\bf 87}, 043401 (2001).

\bibitem{UmrTouFilSorHen-PRL-07}
C.~J. Umrigar, J.~Toulouse, C.~Filippi, S.~Sorella, and R.~G. Hennig,
\newblock Phys. Rev. Lett. {\bf 98}, 110201 (2007).

\bibitem{TouUmr-JCP-07}
J.~Toulouse and C.~J. Umrigar,
\newblock J. Chem. Phys. {\bf 126}, 084102 (2007).

\bibitem{TouUmr-JCP-08}
J.~Toulouse and C.~J. Umrigar,
\newblock J. Chem. Phys. {\bf 128}, 174101 (2008).

\bibitem{Neuscamman:2015:subtractive_jagp}
E.~Neuscamman,
\newblock Mol. Phys. , DOI:10.1080/00268976.2015.1115903 (2015).

\bibitem{POPLE:1953:agp}
A.~C. Hurley, J.~Lennard-Jones, and J.~A. Pople,
\newblock Proc. R. Soc. London, Ser. A {\bf 220}, 446 (1953).

\bibitem{Beran:2005:upp}
G.~J.~O. Beran, B.~Austin, A.~Sodt, and M.~Head-Gordon,
\newblock J. Phys. Chem. A {\bf 109}, 9183 (2005).

\bibitem{Bratoz:1965:AGP}
S.~Brato\v{z} and P.~Durand,
\newblock J. Chem. Phys. {\bf 43}, 2670 (1965).

\bibitem{COLEMAN:1965:AGP}
A.~J. Coleman,
\newblock J. Math. Phys. {\bf 6}, 1425 (1965).

\bibitem{Scuseria:2002:hfb}
V.~N. Staroverov and G.~E. Scuseria,
\newblock J. Chem. Phys. {\bf 117}, 11107 (2002).

\bibitem{Kutzelnigg:1964:apsg}
W.~Kutzelnigg,
\newblock J. Chem. Phys. {\bf 40}, 3640 (1964).

\bibitem{Kutzelnigg:1965:apsg}
W.~Kutzelnigg,
\newblock Theoret. chim. Acta {\bf 3}, 241 (1965).

\bibitem{Surjan:2012:apsg}
P.~R. Surj\'{a}n, \'{A}gnes Szabados, P.~Jeszenszki, and T.~Zoboki,
\newblock J. Math. Chem. {\bf 50}, 534 (2012).

\bibitem{Head-Gordon:2000:non_orth_pp}
T.~V. Voorhis and M.~Head-Gordon,
\newblock J. Chem. Phys. {\bf 112}, 5633 (2000).

\bibitem{Head-Gordon:2000:imperfect_pairing}
T.~V. Voorhis and M.~Head-Gordon,
\newblock Chem. Phys. Lett. {\bf 317}, 575 (2000).

\bibitem{Head-Gordon:2002:gvb_cc}
T.~V. Voorhis and M.~Head-Gordon,
\newblock J. Chem. Phys. {\bf 117}, 9190 (2002).

\bibitem{Bultinck:2013:nonorth_gems}
P.~A. Limacher et~al.,
\newblock J. Chem. Theory Comput. {\bf 9}, 1394 (2013).

\bibitem{Ayers:2014:nonvar_oo_ap1rog}
K.~Boguslawski et~al.,
\newblock J. Chem. Theory Comput. {\bf 10}, 4873 (2014).

\bibitem{VanNeck:2014:ap1rog_on_hubbard}
K.~Boguslawski et~al.,
\newblock Phys. Rev. B {\bf 89}, 201106(R) (2014).

\bibitem{VanNeck:2014:seniority_2_ap1rog_oo}
K.~Boguslawski et~al.,
\newblock J. Chem. Phys. {\bf 140}, 214114 (2014).

\bibitem{Ayers:2014:geminal_accuracy}
P.~Tecmer et~al.,
\newblock J. Phys. Chem. A {\bf 118}, 9058 (2014).

\bibitem{Ayers:2015:ap1rog_lcc}
K.~Boguslawski and P.~W. Ayers,
\newblock J. Chem. Theory Comput. {\bf 11}, 5252 (2015).

\bibitem{Rassolov:2002:ssg}
V.~A. Rassolov,
\newblock J. Chem. Phys. {\bf 117}, 5978 (2002).

\bibitem{Rassolov:2004:pert_ssg}
V.~A. Rassolov, F.~Xu, and S.~Garashchuk,
\newblock J. Chem. Phys. {\bf 120}, 10385 (2004).

\bibitem{Rassolov:2007:ssg}
V.~A. Rassolov and F.~Xu,
\newblock J. Chem. Phys. {\bf 126}, 234112 (2007).

\bibitem{Rassolov:2007:spin_proj_ssg}
V.~A. Rassolov and F.~Xu,
\newblock J. Chem. Phys. {\bf 127}, 044104 (2007).

\bibitem{Rassolov:2014:sspg}
B.~A. Cagg and V.~A. Rassolov,
\newblock J. Chem. Phys. {\bf 141}, 164112 (2014).

\bibitem{Szabados:2015:spin_proj_ssg}
P.~Jeszenszki, P.~R. Surj\'{a}n, and \'{A}gnes Szabados,
\newblock J. Chem. Theory Comput. {\bf 11}, 3096 (2015).

\bibitem{Sorella:2003:agp_sr}
M.~Casula and S.~Sorella,
\newblock J. Chem. Phys. {\bf 119}, 6500 (2003).

\bibitem{Sorella:2004:agp_sr}
M.~Casula, C.~Attaccalite, and S.~Sorella,
\newblock J. Chem. Phys. {\bf 121}, 7110 (2004).

\bibitem{Sorella:2007:jagp_vdw}
S.~Sorella, M.~Casula, and D.~Rocca,
\newblock J. Chem. Phys. {\bf 127}, 014105 (2007).

\bibitem{Sorella:2009:jagp_molec}
M.~Marchi, S.~Azadi, M.~Casula, and S.~Sorella,
\newblock J. Chem. Phys. {\bf 131}, 154116 (2009).

\bibitem{Martinez:2012:thc_correlated}
E.~G. Hohenstein, R.~M. Parrish, C.~D. Sherrill, and T.~J. Martinez,
\newblock J. Chem. Phys. {\bf 137}, 221101 (2012).

\bibitem{Ukrainskii:1977}
I.~I. Ukrainskii,
\newblock Theor. Math. Phys. {\bf 32}, 816 (1977).

\bibitem{Cullen:1996:gvb_from_cc}
J.~Cullen,
\newblock Chem. Phys. {\bf 202}, 217 (1996).

\bibitem{Head-Gordon:2001:gvb_cc}
T.~V. Voorhis and M.~Head-Gordon,
\newblock J. Chem. Phys. {\bf 115}, 7814 (2001).

\bibitem{Scuseria:2014:sen_0_pair_ccd}
T.~Stein, T.~M. Henderson, and G.~E. Scuseria,
\newblock J. Chem. Phys. {\bf 140}, 214113 (2014).

\bibitem{Henderson:2014:pair_ccd_attractive}
T.~M. Henderson, G.~E. Scuseria, J.~Dukelsky, A.~Signoracci, and T.~Duguet,
\newblock Phys. Rev. C {\bf 89}, 054305 (2014).

\bibitem{Scuseria:2014:seniority_cc}
T.~M. Henderson, I.~W. Bulik, T.~Stein, and G.~E. Scuseria,
\newblock J. Chem. Phys. {\bf 141}, 244104 (2014).

\bibitem{Head-Gordon:2000:var_cc}
T.~V. Voorhis and M.~Head-Gordon,
\newblock J. Chem. Phys. {\bf 113}, 8873 (2000).

\bibitem{Knowles:2010:vcc}
B.~Cooper and P.~J. Knowles,
\newblock J. Chem. Phys. {\bf 133}, 234102 (2010).

\bibitem{Knowles:2012:quasi_var_cc}
J.~B. Robinson and P.~J. Knowles,
\newblock J. Chem. Phys. {\bf 136}, 054114 (2012).

\bibitem{Knowles:2012:qvcc_benchmark}
J.~B. Robinson and P.~J. Knowles,
\newblock J. Chem. Theory Comput. {\bf 8}, 2653 (2012).

\bibitem{Knowles:2012:qvcc_nonlin_optical}
J.~B. Robinson and P.~J. Knowles,
\newblock J. Chem. Phys. {\bf 137}, 054301 (2012).

\bibitem{Knowles:2012:qvcc_pert_triples}
J.~B. Robinson and P.~J. Knowles,
\newblock J. Chem. Phys. {\bf 138}, 074104 (2013).

\bibitem{PaldusLi:1999:cc_review}
J.~Paldus and X.~Li,
\newblock Adv. Chem. Phys. {\bf 110}, 1 (1999).

\bibitem{Helgaker_book}
T.~Helgaker, P.~J{\o}rgensen, and J.~Olsen,
\newblock {\em Molecular Electronic Structure Theory},
\newblock John Wiley \& Sons, Ltd., West Sussex, England, 2000.

\bibitem{eigenvalue_templates_2000}
Z.~Bai et~al., editors,
\newblock {\em Templates for the Solution of Algebraic Eigenvalue Problems: A
  Practical Guide},
\newblock SIAM, Philadelphia, 2000.

\bibitem{Davidson:1975:davidson}
E.~R. Davidson,
\newblock J. Comput. Phys. {\bf 17}, 87 (1975).

\bibitem{Arnoldi:1951:arnoldi}
W.~E. Arnoldi,
\newblock Quart. Appl. Math. {\bf 9}, 17 (1951).

\bibitem{Neuscamman:2012:fast_sr}
E.~Neuscamman, C.~J. Umrigar, and G.~K.-L. Chan,
\newblock Phys. Rev. B {\bf 85}, 045103 (2012).

\bibitem{Sorella:2001:SR}
S.~Sorella,
\newblock Phys. Rev. B {\bf 64}, 024512 (2001).

\bibitem{Psi3}
T.~D. Crawford et~al.,
\newblock J. Comput. Chem. {\bf 28}, 1610 (2007).

\bibitem{Werner:1985_1:mcscf}
H.-J. Werner and P.~J. Knowles,
\newblock J. Chem. Phys. {\bf 82}, 5053 (1985).

\bibitem{Werner:1985_2:mcscf}
P.~J. Knowles and H.-J. Werner,
\newblock Chem. Phys. Lett. {\bf 115}, 259 (1985).

\bibitem{Handy:1984:fci}
P.~Knowles and N.~Handy,
\newblock Chem. Phys. Lett. {\bf 111}, 315 (1984).

\bibitem{Handy:1989:fci}
P.~Knowles and N.~Handy,
\newblock Comp. Phys. Commun. {\bf 54}, 75 (1989).

\bibitem{Werner:1996:caspt2}
H.-J. Werner,
\newblock Mol. Phys. {\bf 89}, 645 (1996).

\bibitem{Knowles:1988:mrci}
H.-J. Werner and P.~J. Knowles,
\newblock J. Chem. Phys. {\bf 89}, 5803 (1988).

\bibitem{Werner:1988:mrci}
P.~J. Knowles and H.-J. Werner,
\newblock Chem. Phys. Lett. {\bf 145}, 514 (1988).

\bibitem{MOLPRO_brief}
H.-J. Werner et~al.,
\newblock \uppercase{MOLPRO}, version 2012.1, a package of ab initio programs,
\newblock see http://www.molpro.net.

\bibitem{Szabo-Ostland}
A.~Szabo and N.~S. Ostlund,
\newblock {\em Modern Quantum Chemistry: Introduction to Advanced Electronic
  Structure Theory},
\newblock Dover Publications, Mineola, N.Y., 1996.

\bibitem{QChem:2006}
{Y. Shao et al},
\newblock Phys. Chem. Chem. Phys. {\bf 8}, 3172 (2006).

\bibitem{QChem:2013}
A.~Krylov and P.~Gill,
\newblock WIREs Comput. Mol. Sci. {\bf 3}, 317 (2013).

\bibitem{POPLE:1972:6-31g_basis}
W.~J. Hehre, R.~Ditchfield, and J.~A. Pople,
\newblock J. Chem. Phys. {\bf 56}, 2257 (1972).

\bibitem{Scuseria:2015:ccd0}
I.~W. Bulik, T.~M. Henderson, and G.~E. Scuseria,
\newblock J. Chem. Theory Comput. {\bf 11}, 3171 (2015).

\bibitem{Pople:1977:hf_stability}
R.~Seeger and J.~A. Pople,
\newblock J. Chem. Phys. {\bf 66}, 3045 (1977).

\bibitem{Neuscamman:2009:qct}
E.~Neuscamman, T.~Yanai, and G.~K.-L. Chan,
\newblock J. Chem. Phys. {\bf 130}, 124102 (2009).

\bibitem{FouMitNeeRaj-RMP-01}
W.~M.~C. Foulkes, L.~Mitas, R.~J. Needs, and G.~Rajagopal,
\newblock Rev. Mod. Phys. {\bf 73}, 33 (2001).

\bibitem{Andersson:1995:caspt2_level_shift}
B.~O. Roos and K.~Andersson,
\newblock J. Chem. Phys. {\bf 245}, 215 (1995).

\bibitem{Mavridis:2010:tmo_ScO_TiO_CrO_MnO}
E.~Miliordos and A.~Mavridis,
\newblock J. Phys. Chem. A {\bf 114}, 8536 (2010).

\bibitem{Zhao:2016:evp_qmc}
L.~Zhao and E.~Neuscamman,
\newblock arXiv:1508.06683  (2016).

\end{thebibliography}


\end{document}